\documentclass[range]{ar2e}

\usepackage{amssymb}
\usepackage{amsmath}

\usepackage[dvips]{graphicx}

\usepackage{color}
\definecolor{lightgray}{gray}{0.1}
\definecolor{darkgray}{gray}{0.8}

\usepackage{pstricks,pst-node,pst-text,pst-3d,pst-coil,pst-plot}
\newpsobject{showgrid}{psgrid}{subgriddiv=1,griddots=10,gridlabels=6pt}

\newcommand{\boldvec}[1]{\mbox{\boldmath $#1$}}

\begin{document}
\jname{Annual Review of Nuclear and Particle Science}
\jyear{2004}
\jvol{53}
\ARinfo{1056-8700/97/0610-00}

\title{Heavy Quarks on the Lattice}

\markboth{Hashimoto \& Onogi}{Heavy Quarks on the Lattice}

\author{
  Shoji Hashimoto
  \affiliation{
    Institute of Particle and Nuclear Studies,\\
    High Energy Accelerator Research Organization (KEK),\\
    Tsukuba 305-0801, JAPAN
  }
  Tetsuya Onogi
  \affiliation{
    Yukawa Institute for Theoretical Physics, 
    Kyoto University,\\
    Kyoto 606-8502, JAPAN
  }
}

\begin{keywords}
lattice QCD, quarkonium, $B$ meson
\end{keywords}

\begin{abstract}
  Lattice quantum chromodynamics provides first principles
  calculations for hadrons containing heavy quarks ---
  charm and bottom quarks.
  Their mass spectra, decay rates, and some hadronic matrix 
  elements can be calculated on the lattice in a model
  independent manner.
  In this review, we introduce the effective theories
  that treat heavy quarks on the lattice.
  We summarize results on the heavy quarkonium spectrum, which
  verify the validity of the effective theory approach.
  We then discuss applications to $B$ physics, which
  is the main target of the lattice theory of heavy quarks.
  We review progress in lattice calculations of
  the $B$ meson decay constant, the $B$ parameter,
  semi-leptonic decay form factors, and other important 
  quantities.
\end{abstract}

\maketitle

\section{Introduction}
\label{sec:Introduction}

Quantum chromodynamics (QCD) is the fundamental theory
of the strong interaction.
It is the SU(3) gauge field theory, which describes the
interaction among quarks carrying three (R, G, and B)
internal degrees of freedom, conveniently called
``color.''
Because of the non-Abelian nature of gauge symmetry, the
gluon (the gauge field in QCD) interacts with itself, and as
a result the strong coupling constant $\alpha_s$ becomes
large as the momentum of the exchanged gluon decreases.
Therefore, the calculation of low energy (approximately
several hundred MeV) properties of hadrons, such as their
masses and interactions, is a challenging problem that
requires nonperturbative methods. 

A regularized formulation of QCD on a four-dimensional
hypercubic lattice, called lattice QCD
\cite{Wilson:1974sk}, enables us to calculate such hadron
properties in the strong coupling regime. 
Because the interaction is highly nonlinear and the number
of degrees of freedom is huge (proportional to the
space-time volume), the lattice calculation is
computationally so demanding that several dedicated
supercomputers have been developed around the world.
The application of lattice QCD covers a wide area including 
the light hadron mass spectrum, finite temperature phase
transition, and weak-interaction matrix elements --- as well
as heavy quark physics, which is the subject of this article.

Among the six flavors of quarks in the Standard Model,
charm ($c$), bottom ($b$), and top ($t$) quarks are heavy
compared to the hadronic energy scale.
The top quark decays very rapidly to a real $W$ boson and
bottom quark, and its QCD interaction can be treated
perturbatively. 
Therefore, we consider the physics of charm and bottom
quarks.  

The main motivation for charm and bottom quark physics
is to study the flavor structure of the Standard Model
including the mechanism that induces CP violation.
Detailed study of the decays of charmed and bottom mesons
provides pieces of information to precisely determine the
Cabibbo-Kobayashi-Maskawa (CKM) matrix elements, which are 
fundamental parameters in the Standard Model.
Furthermore, the loop effect of physics beyond the Standard
Model may be probed through flavor-changing neutral
current (FCNC) processes, such as $B^0$-$\bar{B}^0$ 
($D^0$-$\bar{D}^0$) mixing or rare $B$ decays, which are
suppressed in the Standard Model by the 
Glashow-Iliopuolos-Maiani (GIM) mechanism
\cite{Glashow:gm}. 
In such studies model-independent calculations of the
hadronic matrix elements are crucial in order to precisely
predict the Standard Model contributions.

One example is the rate (or frequency) of the neutral $B$
meson mixing, which is induced by $\Delta B=2$ box
diagrams. 
Because of the GIM mechanism, the dominant contribution
comes from the diagram involving the top quark, which is
proportional to the CKM matrix element $|V_{td}|$ squared.
At the $B$ meson energy scale, this interaction is
effectively described by a local operator 
$O^{\Delta B=2} =
\bar{b}\gamma_\mu(1-\gamma_5)q\,
\bar{b}\gamma_\mu(1-\gamma_5)q$,
and its rate is proportional to a matrix element
$\langle \bar{B}|O^{\Delta B=2}|B\rangle$, which is
parametrized by the $B$ meson decay constant $f_B$ and the
$B$-parameter $B_B$.
This matrix element represents the probability of finding
the heavy quark and the light anti-quark at the same spatial 
point in the $B$ meson to annihilate it and to create a
$\bar{B}$ meson from that point.
This is a problem of bound state formation in QCD, which is
non-perturbative.
The lattice calculation offers the best tool to solve such
problems, but one needs a special treatment of heavy quark
fields on the lattice, since the Compton wavelength of
the heavy quark, $\sim 1/m_Q$, could be smaller than the
typical lattice spacing $a$.

The typical energy or momentum scale that governs the
dynamics of quarks and gluons inside the usual light hadrons
is the QCD scale $\Lambda_{\mathrm{QCD}}$, which
characterizes the energy scale where the QCD coupling
$\alpha_s$ becomes large.
Because the heavy quark introduces a new energy scale to the
system, the QCD dynamics inside heavy hadrons is
quite different from that of light hadrons.

\begin{figure}[t]
  \centering
  \begin{pspicture}[1.0](0.0,-0.2)(6.0,6.0)
    \multido{\ny=1+0.5}{9}{\psline(1,\ny)(5,\ny)}
    \multido{\nx=1+0.5}{9}{\psline(\nx,1)(\nx,5)}
    \pscircle[linestyle=none,fillstyle=solid,fillcolor=lightgray](3,3){1.7}
    \rput[B](3.6,2.2){\large $\bar{q}$, $g$}
    \pscircle[fillstyle=solid,fillcolor=darkgray](3,3){0.2}
    \rput[B](2.55,2.55){\large $Q$}
    \psline{<->}(2.8,3.3)(3.2,3.3)
    \rput[B](3.1,3.6){$m_Q^{-1}$}
    \psline{<->}(1,0.8)(1.5,0.8)
    \rput[B](1.25,0.5){$a$}
    \psline{<->}(1,0.3)(5,0.3)
    \rput[B](3,-0.1){$L$}
    \psline{<->}(1.6,5.2)(4.4,5.2)
    \rput[B](3,5.4){$\Lambda_{\mathrm{QCD}}^{-1}$}
  \end{pspicture}
  \begin{pspicture}[1.0](0.0,-0.2)(6.0,6.0)
    \multido{\ny=1+0.5}{9}{\psline(1,\ny)(5,\ny)}
    \multido{\nx=1+0.5}{9}{\psline(\nx,1)(\nx,5)}
    \pscircle[linestyle=none,fillstyle=solid,fillcolor=lightgray](3,3){1.2}
    \pscircle[fillstyle=solid,fillcolor=darkgray](3.6,3){0.2}
    \rput[B](3.9,2.55){\large $\bar{Q}$}
    \psarc[linewidth=1.2pt]{<-}(3,3){0.6}{290}{360}
    \pscircle[fillstyle=solid,fillcolor=darkgray](2.4,3){0.2}
    \rput[B](2.1,2.55){\large $Q$}
    \psarc[linewidth=1.2pt]{<-}(3,3){0.6}{110}{180}
    \psline{<->}(3.4,3.3)(3.8,3.3)
    \rput[B](3.7,3.6){$m_Q^{-1}$}
    \psline{<->}(1,0.8)(1.5,0.8)
    \rput[B](1.25,0.5){$a$}
    \psline{<->}(1,0.3)(5,0.3)
    \rput[B](3,-0.1){$L$}
    \psline{<->}(2.1,5.2)(3.9,5.2)
    \rput[B](3,5.4){$(m_Qv)^{-1}$}
  \end{pspicture}
  \caption{
    Typical momentum scales in the heavy-light (left) and
    heavy-heavy (right) mesons.
  } 
  \label{fig:scales}
\end{figure}
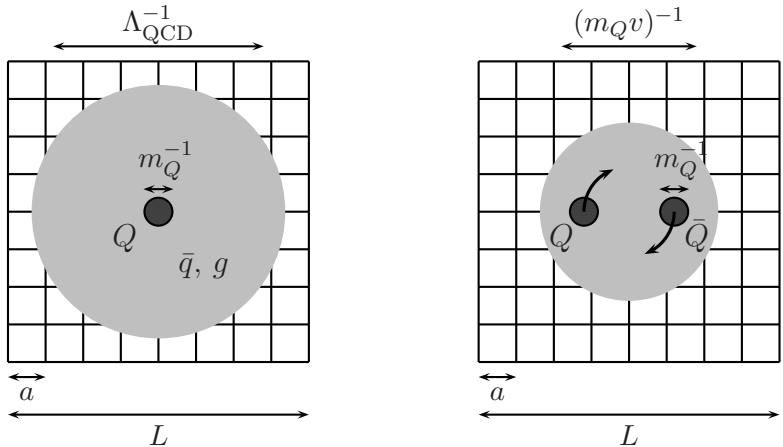

In the heavy-light meson, which is a bound state composed 
of a heavy quark $Q$ and light degrees of freedom
(a light anti-quark $\bar{q}$ and gluons $g$), 
the motion of the heavy quark of mass $m_Q$ is hardly
affected by the light degrees of freedom with a typical
momentum $\Lambda_{\mathrm{QCD}}$, if
$m_Q\gg\Lambda_{\mathrm{QCD}}$.
Thus, the heavy quark stays almost at rest at the center of
the bound state, surrounded by the light degrees of freedom,
as shown in the left panel of Figure~\ref{fig:scales}.
The motion of the heavy quark is suppressed by
$\Lambda_{\mathrm{QCD}}/m_Q$.
This system is described by the Heavy Quark
Effective Theory (HQET)
\cite{Eichten:1989zv,Georgi:1990um,Grinstein:1990mj}, which
is discussed in Section~\ref{sec:HQET_and_NRQCD}.

The dynamics of quarkonium, which is a bound state of 
$Q$ and  $\bar{Q}$, is governed by different energy scales. 
In the classical picture, the nonrelativistic kinetic energy
$\langle p^2\rangle/2m_Q$ and the potential energy 
$-\frac{4}{3}\alpha_s\langle\frac{1}{r}\rangle$ have to be
balanced, and the heavy quarks move around each other,
in contrast to the heavy-light dynamics, as depicted in
Figure~\ref{fig:scales} (right panel).
$\langle p\rangle$ and $\langle\frac{1}{r}\rangle$ denote
the typical size of the relative spatial momentum and
distance between the two heavy quarks, respectively.
The uncertainty relation 
$\langle p\rangle \sim \langle 1/r\rangle$ implies
$\langle p\rangle \sim \alpha_s m_Q$, 
which means that the typical velocity 
$v\sim \langle p\rangle/m_Q$ is of the order of 
the strong coupling constant $\alpha_s$.
Thus, there are three momentum scales in the heavy
quarkonia:
heavy quark mass $m_Q$, 
typical spatial momentum $\langle p\rangle \simeq m_Qv$, 
and typical binding energy 
$\langle p^2\rangle/m_Q \simeq m_Q v^2$. 
It is possible to construcnt an effective theory that
explicitly separates these energy scales.
This is known as non-relativistic QCD (NRQCD)
\cite{Caswell:1985ui,Thacker:1990bm,Lepage:1992tx}.

A number of lattice simulations for heavy quarks have been 
performed with or without these effective theories.
For the heavy-light mesons, the simplest and best known
application is the calculation of the $B$ meson decay
constant. 
Other more complicated quantities include the $B$-parameters 
and the semi-leptonic decay form factors.
These quantities are not known from experimental data,
and are necessary inputs to determine the CKM matrix
elements. 
For the heavy-heavy mesons, on the other hand, the mass
spectrum of $c\bar{c}$ and $b\bar{b}$ quarkonia is
experimentally very well known, and thus provides a
benchmark by which to test the validity of the lattice 
calculations.

Until recently, because of the lack of computer power, most
lattice simulations have been done in the quenched
approximation, in which the effect of quark-antiquark
pair creation (and annihilation) in the vacuum is
neglected. 
This approach could introduce uncontrolled systematic
uncertainty, and hence the results should not be taken as a
model independent prediction.
However, the full QCD simulation without such an
approximation has become realistic recently.
We review some results from such calculations.

This article is organized as follows.
In Section~\ref{sec:HQET_and_NRQCD} we introduce the
effective theories used for heavy-light and heavy-heavy
systems, and discuss their lattice formulations.
Matching of the effective theory onto the continuum QCD is
one of the key issues.
Some results for quarkonia mass spectrum and decays are
discussed in Section~\ref{sec:Quarkonia}.
Model independent calculation of the matrix elements
required in $B$ physics phenomenology is the main
motivation for the study of heavy quark on the lattice.
We review the current status in 
Section~\ref{sec:B_physics}. 
Finally, the prospects for future lattice calculations
are discussed in Section~\ref{sec:Perspectives}.

We do not discuss the determination of the bottom quark mass
using the lattice results for quarkoinum because it was well
covered in a recent review by El-Khadra \& Luke
\cite{El-Khadra:2002wp}. 
Reviews presented at recent lattice conferences 
\cite{Bernard:2000ki,Ryan:2001ej,Yamada:2002wh,Kronfeld:2003sd}
and in other review articles
\cite{Flynn:1997ca,Kronfeld:2003sd}
are also useful sources of information.
For textbooks on lattice gauge theory, see for example
\cite{Montvay:cy,Smit:ug}.


\section{HQET and NRQCD}
\label{sec:HQET_and_NRQCD}

The lattice spacing $a$ in today's typical lattice
simulations is about 0.1--0.2~fm.
The available physical volume is then
$V\simeq (2\mathrm{~fm})^3$ for the typical lattice size
$L/a=$ 16--24, which is numerically feasible.
Because the Compton wavelength of charm and bottom quarks 
($1/m_c\simeq 0.13\mathrm{~fm}$ and 
$1/m_b\simeq 0.04\mathrm{~fm}$, respectively)
are similar to or even smaller than the lattice spacing,
discretization errors would seem to be out of control.
Naively, the discretization error grows as $(am_Q)^n$,
and the power $n$ is 2 for the $O(a)$-improved lattice
fermion actions, such as the  $O(a)$-improved Wilson fermion 
\cite{Sheikholeslami:1985ij}. 

However, the non-perturbative dynamics of QCD becomes
important only in the low energy
($\sim\Lambda_{\mathrm{QCD}}$) regime.
Therefore, if one can formulate the theory such that the 
energy scale of order $m_Q$ is explicitly separated from
the low energy degrees of freedom, it would suffice to treat 
only the low energy part in the lattice simulation; the high
energy part can be reliably treated in perturbation theory.
Theoretically, such separation of different energy scales
can be formulated in terms of the 
Operator Product Expansion (OPE) \cite{Wilson:zs}.
Its well-known application is in deep inelastic scattering,
for which the large energy scale comes the energy of
collisions.
Lagrangian-based formulations are more transparent in many
applications, and for the heavy-light and heavy-heavy
hadrons dedicated formulations have been developed:
Heavy Quark Effective Theory (HQET) for the heavy-light and 
nonrelativistic QCD (NRQCD) for the heavy-heavy hadrons.

\subsection{Continuum HQET and NRQCD}
Let us consider a heavy hadron in its rest frame.
The momentum of the heavy quark inside the heavy hadron can
be written as $p_Q^\mu=(m_Q+k^0,\boldvec{k})$ by separating
the heavy quark mass $m_Q$ from other, smaller momentum.
From the heavy quark field $q$ one may extract the trivial
dependence on the heavy quark mass as
$q = e^{-im_Qt}
\left(\begin{array}[c]{c}Q\\X\end{array}\right)$,
where the four-component spinor is divided into two
two-component spinors $Q$ and $X$.
Then, the Dirac equation 
$(i\gamma^\mu D_\mu - m_Q) q = 0$
with the Dirac representation of the $\gamma$ matrices
can be separated into two parts:
\begin{eqnarray}
  iD_0\, Q & = & i\boldvec{\sigma}\cdot\boldvec{D}\, X,
  \label{eq:Dirac_eq_upper}
  \\
  (2m_Q+ iD_0)\, X & = & i\boldvec{\sigma}\cdot\boldvec{D}\, Q,
  \label{eq:Dirac_eq_lower}
\end{eqnarray}
where $\sigma^i$ are the Pauli matrices.
The second equation shows that the lower component $X$
is smaller than the upper component $Q$ by a factor $2m_Q$.
If we can neglect the small time-dependence of $X$, $iD_0 X$
in Equation~\ref{eq:Dirac_eq_lower}, and substitute 
$X=i\boldvec{\sigma}\cdot\boldvec{D}/2m_Q Q$ 
into Equation~\ref{eq:Dirac_eq_upper}, we arrive at the
non-relativistic Shr\"odinger equation plus the Pauli term,
\begin{equation}
  \label{eq:Pauli_equation}
  iD_0\, Q = - \left[
    \frac{\boldvec{D}^2}{2m_Q} +
    \frac{\boldvec{\sigma}\cdot g\boldvec{B}}{2m_Q}
  \right] Q,
\end{equation}
where $g$ is the QCD coupling constant and
$B^i=\frac{1}{2}\epsilon^{ijk}F^{jk}$ is the
magnetic part of the field-strength tensor of QCD.
In Lagrangian form we may write this as
\begin{eqnarray}
  \label{eq:L_heavy}
  \mathcal{L}_{\mathrm{heavy}} 
  & = &
  Q^\dagger \left[
    iD_0 + \frac{\boldvec{D}^2}{2m_Q} +
    \frac{\boldvec{\sigma}\cdot g\boldvec{B}}{2m_Q}
    + \frac{\boldvec{D}\cdot g\boldvec{E}-g\boldvec{E}\cdot\boldvec{D}}{8m_Q^2}
  \right.
  \nonumber\\
  & &
  \left.
    + \frac{\boldvec{\sigma}\cdot(
      i\boldvec{D}\times g\boldvec{E}-g\boldvec{E}\times i\boldvec{D})}{8m_Q^2}
    + \frac{(\boldvec{D}^2)^2}{8m_Q^3}
    + \cdots
  \right] Q,
\end{eqnarray}
recovering some of the higher order contributions.
The ellipses represent neglected higher order terms.

In the heavy-light hadron, the light degrees of freedom
(light quarks and gluons) have momenta of order
$\Lambda_{\mathrm{QCD}}$.
Exchange of spatial momenta with the heavy quark occurs
through the $1/m_Q$ and higher order terms in
Equation~\ref{eq:L_heavy}.
Thus, the motion of the heavy quark is suppressed by powers
of $\Lambda_{\mathrm{QCD}}/m_Q$.
In HQET 
\cite{Eichten:1989zv,Georgi:1990um,Grinstein:1990mj}
one treats the Lagrangian
equation~(Equation~\ref{eq:L_heavy}) as a systematic
expansion in $1/m_Q$. 
At leading order, only the first term $Q^\dagger iD_0 Q$
remains, which represents a non-moving heavy quark acting
only as a static color source.
This is often called the static approximation.
The higher order terms can be included by operator
insertions, that is, by expanding 
$\exp(i\int d^4x\mathcal{L}_{heavy})$
in terms of $1/m_Q$.
In the lattice simulations, it is more convenient to 
include the higher order terms as a part of the heavy quark
propagator.

In the heavy-heavy hadron, on the other hand, the momentum 
of the heavy (anti-)quark is determined by the balance
between the potential energy and kinetic energy, as already 
mentioned in Section~\ref{sec:Introduction}.
Specifically, in order to satisfy
Equation~\ref{eq:Pauli_equation},
$\langle iD_0 Q\rangle \sim 
 \langle \boldvec{D}^2/2m_Q Q\rangle$.
Because the heavy quark potential is well described by the
Coulomb form $-C_F\alpha_s/r$, this implies
\begin{equation}
  \langle\frac{\alpha_s}{r}\rangle
  \sim
  \langle\alpha_s p\rangle
  \sim
  \frac{\langle p^2 \rangle}{m_Q}.
\end{equation}
Then, the typical momentum is of the order $m_Q\alpha_s$, or
the typical velocity $v=\langle p\rangle/m_Q$ is of order
$\alpha_s$. 
The separation of the energy scales is provided by the
velocity $v$: 
(heavy quark mass $m_Q$) $\gg$ 
(momentum $m_Qv$) $\gg$ (binding energy $m_Qv^2$).

In NRQCD 
\cite{Caswell:1985ui,Thacker:1990bm,Lepage:1992tx}
the counting of terms in the Lagrangian
equation (Equation~\ref{eq:L_heavy}) is done by the power of
$v$. 
Each operator has a different power counting:
$Q\sim (m_Qv)^{3/2}$,
$\boldvec{D}\sim m_Qv$, 
$D_0\sim m_Qv^2$, 
$g\boldvec{E}\sim m_Q^2v^3$,
and $g\boldvec{B}\sim m_Q^2v^4$.
As a result, the leading terms in Equation~\ref{eq:L_heavy}
are $Q^\dagger iD_0Q$ and 
$Q^\dagger(-\boldvec{D}^2/2m_Q)\,Q$;
the other terms are suppressed by a relative power $v^2$.
Higher order terms can also be included in a systematic
manner. 

The QCD coupling constant $\alpha_s$ depends on the energy
scale, according to the renormalization group running.
It should be evaluated at the momentum of the exchanged
gluons, which in this case is the momentum scale of
$m_Qv\sim m_Q\alpha_s$.
By solving $v\sim\alpha_s(m_Qv)$ self-consistently, the
typical velocity squared is estimated to be $v^2\sim$ 0.3
for charmonium ($c\bar{c}$) and 0.1 for bottomonium
($b\bar{b}$). 
Expansion in $v^2$ is therefore very effective for
$b\bar{b}$ but marginal for $c\bar{c}$.

\subsection{Lattice HQET and NRQCD}
Once the effective theories are introduced, the length scale
to be treated on the lattice is not $1/m_Q$, but
$1/\Lambda_{\mathrm{QCD}}$ for heavy-light or $1/m_Qv$ for
heavy-heavy systems. 
These are long enough to be approximated by a lattice
of $a\sim$~0.1--0.2~fm.
A rough sketch of the relevant length scales is shown
in Figure~\ref{fig:scales}.

Discretization of the Lagrangian equation
(Equation~\ref{eq:L_heavy}) is straightforward.
First, one must perform the Wick rotation $x^0=ix_4$ to
move on to the Euclidean space-time.
Then, the covariant derivative $D_\mu$ is replaced by a
finite difference
\begin{equation}
  \label{eq:lattice_covariant_derivative+}
  \Delta^{(+)}_\mu Q(x) \equiv
  U_\mu(x) Q(x+\hat{\mu}) - Q(x),
\end{equation}
where $x+\hat{\mu}$ denotes a lattice point next to $x$ in
the $\mu$-direction.
$U_\mu(x)$ is the link variable, which represents
the gauge field on the lattice:
the relation to the continuum vector potential $A_\mu(x)$ is 
approximately given by $U_\mu(x)=\exp[iagA_\mu(x)]$.
The discretized derivative,
Equation~\ref{eq:lattice_covariant_derivative+}, is
constructed such that it is covariant under the SU(3) gauge 
transformation. 
One can also define
\begin{equation}
  \label{eq:lattice_covariant_derivative-}
  \Delta^{(-)}_\mu Q(x) \equiv
  Q(x) - U^\dagger_\mu(x-\hat{\mu}) Q(x-\hat{\mu}),
\end{equation}
$\Delta^{(\pm)}_\mu = (\Delta^{(+)}+\Delta^{(-)})/2$,
and the lattice Laplacian 
$\Delta^{(2)}\equiv\sum_i\Delta_i^{(+)}\Delta_i^{(-)}$.

One explicit form of the lattice action is
\begin{equation}
  \label{eq:lattice_action}
  S = \sum_{t,\boldvec{x}}
  Q^\dagger(t,\boldvec{x})
  [Q(t,\boldvec{x})-K_t Q(t-1,\boldvec{x})],
\end{equation}
where $t=x_4$.
The operator $K_t$ is a kernel of time-evolution:
\begin{equation}
  \label{eq:evolution_kernel}
  K_t =
  \left(1-\frac{aH_0}{2n}\right)_t^n
  \left(1-\frac{a\delta H}{2}\right)_t
  U_{4,t-1}^\dagger
  \left(1-\frac{a\delta H}{2}\right)_{t-1}
  \left(1-\frac{aH_0}{2n}\right)_{t-1}^n.
\end{equation}
Here subscripts represent the time slice at which
Hamiltonian operators such as $(1-aH_0/2n)$ act, and an
integer $n$ is introduced to suppress instability that
appears in the evolution equation owing to unphysical
momentum modes \cite{Thacker:1990bm,Lepage:1992tx}.
The leading order Hamiltonian $H_0$ is given by
\begin{equation}
  \label{eq:H_0}
  H_0 = - \frac{\Delta^{(2)}}{2m_Q},
\end{equation}
and the higher order terms are 
\begin{eqnarray}
  \label{eq:delta_H}
  \delta H & = &
  - \frac{\boldvec{\sigma}\cdot g\boldvec{B}}{2m_Q}
  - \frac{\boldvec{\Delta}^{(\pm)}\cdot g\boldvec{E}
    -g\boldvec{E}\cdot\boldvec{\Delta}^{(\pm)}}{8m_Q^2}
  - \frac{\boldvec{\sigma}\cdot(
    \boldvec{\Delta}^{(\pm)}\times g\boldvec{E}
    -g\boldvec{E}\times\boldvec{\Delta}^{(\pm)})}{8m_Q^2}
  \nonumber\\
  & &
  - \frac{(\Delta^{(2)})^2}{8m_Q^3}
  + \cdots.
\end{eqnarray}
The field strength tensors $\boldvec{B}$ and $\boldvec{E}$
are made from link variables connected on a plaquette. 
Details of their construction are given, for instance, in
Reference~\cite{Lepage:1992tx}.
One can also define other heavy quark actions, which are
equivalent to Equation~\ref{eq:lattice_action} up to
discretization effects.

Because the propagation of the heavy quark is simply
expressed by the time-evolution kernel,
Equation~\ref{eq:evolution_kernel},
numerical calculation of the heavy quark propagator is much 
faster than calculation of the light quark propagator, which
requires some iterative solver, such as the conjugate
gradient method. 

As in the continuum case, the difference between HQET and
NRQCD is in the underlying dynamics.
We can use the same lattice action, but the size of the 
contribution from each term in the Lagrangian differs 
between heavy-light and heavy-heavy hadrons.
In other words, at a given accuracy (say
$\Lambda_{\mathrm{QCD}}/m_Q$ in the heavy-light, or $v^2$ in
the heavy-heavy), the terms to be included in the
calculation are different.

To recover the original four-component Dirac spinor $\psi_h$
from the two-component fields, one must apply the
Foldy-Wouthuysen-Tani (FWT) transformation 
\begin{equation}
  \label{eq:FWT}
  \psi_h =
  \left[
    1-\frac{\boldvec{\gamma}\cdot\Delta^{(\pm)}}{2m_Q}
    +\frac{\Delta^{(2)}}{8m_Q^2}
    +\frac{\boldvec{\Sigma}\cdot g\boldvec{B}}{8m_Q^2}
    -\frac{i\gamma_4\boldvec{\gamma}\cdot g\boldvec{E}}{4m_Q^2}
    + \cdots
  \right]
  \left(\begin{array}[c]{c} Q\\0 \end{array}\right),
\end{equation}
with $\Sigma^j=\mathrm{diag}\{\sigma^j,\sigma^j\}$.
It is necessary when we construct operators that involve
both heavy and light quark fields, such as the heavy-light 
axial-vector current
$A_\mu=\bar{\psi}_h\gamma_5\gamma_\mu\psi_l$. 

In the language of the functional integral to define the
quantum field theory, the FWT transformation is nothing but
a change of variables, and the physics described by the
theories before and after the transformation is the same up
to neglected higher order terms in Equation~\ref{eq:FWT}.
There is a class of possible such transformations, but the
choice in Equation~\ref{eq:FWT} is special because it
removes the lower components from the theory.
This suggests that there are other effective theories
for which the FWT transformation is partially applied.
Such a formulation is known as the Fermilab action
\cite{El-Khadra:1996mp,Kronfeld:2000ck}.
It uses the usual four-component spinors on the lattice, and
the higher spatial derivatives are introduced according to
the order counting of HQET or NRQCD.
This formulation is useful, because at the leading order (of
$1/m_Q$ or $v^2$) the action can be taken to be the same as
that of light quark actions, \textit{i.e.} the Wilson
fermion with or without the $O(a)$-improvement.
Therefore, this formalism covers entire mass regions (from
light to heavy) with a single fermion action, although the
achievable accuracy varies depending on the quark mass.
In the heavy quark regime the accuracy is estimated in terms
of the HQET or NRQCD counting rule.
The discretization effect can also be estimated by
interpreting the lattice effective action with the Symanzik
effective theory \cite{Kronfeld:2000ck}.
To correctly incorporate the higher order corrections one
must introduce higher derivative terms to the action just
as in the NRQCD action.
An equivalent formulation, but without the FWT
transformation, was proposed later in a different context 
\cite{Aoki:2001ra}.

\subsection{Renormalization and continuum limit of the
  effective theories} 
Although the HQET and NRQCD capture the physical picture of
the heavy quark inside the heavy hadrons, the resulting field
theory is a non-renormalizable effective theory.
This means that an infinite number of terms are needed in
order to eliminate ultraviolet divergences that appear in 
quantum loop calculations of the theory.
In fact the Lagrangian equation (Equation~\ref{eq:L_heavy})
contains infinitely many terms at higher orders, and to
renormalize them one needs an infinite number of
renormalization conditions, 
\textit{i.e.} input parameters.
In practice, one is only interested in finite orders of
$1/m_Q$ (or $v^2$) and truncates the Lagrangian at that
order.
Then, the number of renormalization conditions is finite and
the calculation is feasible.
This is how these effective theories work: at a certain
order of the systematic expansion the number of parameters
is still finite, and thus the predictive power remains.

On the lattice with a finite lattice spacing no divergence
appears, but we still need input parameters.
The coefficients in front of the terms in
Equation~\ref{eq:delta_H} are correct only at tree level;
if we include the quantum effects, they become
renormalized. 
These input parameters are provided by calculating some
physical amplitudes in the full relativistic theory.
The coefficients are then determined such that the effective
theory gives the same physical amplitude at a given order of
$1/m_Q$ (or $v^2$).
This procedure is called matching, and it is usually done in
perturbation theory (perturbative matching).

At leading order the necessary terms are $Q^\dagger Q$,
$Q^\dagger D_0Q$ and $Q^\dagger\boldvec{D}^2Q$.
Their renormalization corresponds to the energy shift, wave
function renormalization, and mass renormalization,
respectively. 
The perturbative matching of these parameters onto continuum
full theory was calculated for several specific definitions
of lattice NRQCD actions 
\cite{Davies:1991py,Morningstar:1993de,Morningstar:1994qe,%
  Ishikawa:1999xu}
and for the Fermilab action \cite{Mertens:1997wx},
although the complete calculation of higher order terms,
especially the term $\boldvec{\sigma}\cdot\boldvec{B}$, is
still to be done.

Similarly, the operators to be measured on the lattice 
must be matched onto the continuum full theory.
For example, the heavy-light axial-vector current
$A_\mu=\bar{\psi}_h\gamma_5\gamma_\mu\psi_l$, which is
related to the $B$ meson leptonic decay constant, is
renormalized as
\begin{equation}
  \label{eq:A_4}
  \mathcal{A}_4 = Z_A \left[
    \bar{\psi}_h\gamma_5\gamma_4\psi_l
    - \frac{c_A^{(1)}}{2m_Q}
    \bar{\psi}_h\gamma_5\gamma_4
    \boldvec{\gamma}\cdot\boldvec{\Delta}^{(\pm)}\psi_l
    + \frac{c_A^{(2)}}{2m_Q}
    \bar{\psi}_h\gamma_5\gamma_4
    \boldvec{\gamma}\cdot\overleftarrow{\boldvec{\Delta}}^{(\pm)}\psi_l
  \right],
\end{equation}
for the continuum heavy-light axial-vector current
$\mathcal{A}_\mu$ defined, for instance, using the
$\overline{\mathrm{MS}}$ scheme.
The matching parameters $Z_A$, $c_A^{(1)}$, and $c_A^{(2)}$ 
are calculated at the one-loop level of perturbation
theory. 
In the static limit ($m_Q\rightarrow\infty$), the
perturbative calculation of $Z_A$ was carried out for both
the Wilson
\cite{Boucaud:1989ga,Eichten:1989kb,Boucaud:1992nf} and the
$O(a)$-improved Wilson 
\cite{Borrelli:1992fy,Hernandez:1992su} fermions for the
light quark, and even a non-perturbative calculation has
been done recently \cite{Heitger:2003xg}.
One-loop calculations are also available for the NRQCD
\cite{Davies:ec,Morningstar:1997ep,Morningstar:1998yx,%
  Ishikawa:1999xu} 
and for the Fermilab \cite{Harada:2001fi} lattice actions.
The matching of the four-quark operator 
$O_L=
\bar{\psi}_h\gamma_5(1-\gamma_\mu)\psi_l
\bar{\psi}_h\gamma_5(1-\gamma_\mu)\psi_l$,
which is needed for the the $B^0$-$\bar{B}^0$ mixing matrix 
elements, has also been calculated 
\cite{Flynn:1990qz,Borrelli:1992fy,DiPierro:1998ty,%
  Gimenez:1998mw,Ishikawa:1998rv,Hashimoto:2000eh}.

The continuum limit of the lattice HQET/NRQCD is not
trivial. 
Beyond the static approximation, the lattice action contains 
higher dimensional operators that induce power divergent
terms such as $1/(am_Q)^n$ in the radiative corrections.
Because the divergences are unphysical and must be
subtracted (or renormalized), the perturbative matching
becomes less accurate as $a\rightarrow 0$, even though the
physical picture of the $\Lambda_{\mathrm{QCD}}/m_Q$ 
(or $v^2$) expansion does not change. 
Therefore, the lattice calculation is practically restricted
to relatively large lattice spacings, such that 
$am_Q\gtrsim 1$ is maintained.
This means that the continuum limit $a\rightarrow 0$ cannot
be reached within the effective theories beyond the leading
order (static limit), and one must check the stability of
the results in a limited region of lattice spacings to keep 
$am_Q\gtrsim 1$.
For the Fermilab action \cite{El-Khadra:1996mp}, on the
other hand, the continuum limit can smoothly be reached,
because it is reduced to the Wilson-like fermions as
$am_Q\rightarrow 0$.

In the dimensional regularization adopted in the continuum
perturbation theory, the problem of power divergences in
the OPE appears as the renormalon ambiguity
\cite{Martinelli:1996pk}; that is, 
the lowest order Wilson coefficient contains an ambiguity of 
order $\Lambda_{\mathrm{QCD}}/m_Q$ due to the poor
convergence behavior of the perturbative expansion.
The renormalon ambiguity cancels with the higher order
matrix element of the same order
($\Lambda_{\mathrm{QCD}}/m_Q$), and the physical quantity to 
be computed is free from the ambiguity.
In the theory with an explicit cutoff, such as lattice gauge 
theory, the problem becomes the difficulty of subtracting
unphysical power divergences using the perturbative
expansion, as discussed above.
In practice, the perturbative expansion converges well 
as far as one keeps $am_Q\gtrsim 1$, 
but higher order perturbative calculations are needed to
obtain better control of systematic errors
\cite{Bernard:2000ki,Kronfeld:2003sd}.

One can avoid the whole problem by performing the matching
calculation non-perturbatively, as recently proposed by the
ALPHA collaboration \cite{Heitger:2003nj}.
The matching of lattice HQET onto the relativistic action is 
feasible on a lattice of physically small volume, $L\simeq$
0.2~fm, because one can work with a small lattice spacing for
which the relativistic action can be safely used.
The matching of HQET on larger volumes is then performed
by comparing the lattices of size $L$ and $2L$ recursively.
At present, the method is applied to the leading-order HQET
(static limit), but if the matching could be done in a
similar manner for higher dimensional operators, it would 
provide a breakthrough to go beyond perturbative matching.

\subsection{Relativistic lattice actions used for heavy
  quarks} 
In the usual lattice fermion actions, the quark mass is
included as a perturbation to the massless limit.
Because the theory is renormalizable, the number of relevant
operators is limited and the transition toward the
continuum limit is straightforward.

However, in practical simulations with present computer 
resources, the lattice spacing $a$ is not always small
enough to satisfy $am_Q\ll 1$ for the charm quark, and it is 
absolutely too large for the bottom quark.
The discretization error starts from $O(am_Q)$ for the
unimproved Wilson fermion, or from $O[(am_Q)^2]$ 
for the $O(a)$-improved actions.

One can see how the discretization errors behave by taking
the energy-momentum dispersion relation as an example.
The non-relativistic effective Hamiltonian of the
``relativistic'' lattice fermion can be written as
\begin{equation}
  \mathcal{H}_{\mathrm{heavy}} =
  Q^\dagger \left[
    m_1 - \frac{\boldvec{D}^2}{2m_2} + \cdots
  \right] Q,
\end{equation}
where the rest mass $m_1$ and the kinetic mass $m_2$ can be
different if the Lorentz invariance is lost owing to the
discretization error.
At tree level they become
\begin{eqnarray}
  \label{eq:m1}
  m_1a & = & \ln(1+m_0a), 
  \\
  \label{eq:m2}
  \frac{1}{m_2a} & = &
  \frac{2}{m_0a(2+m_0a)} + \frac{1}{1+m_0a},
\end{eqnarray}
for the $O(a)$-improved and -unimproved Wilson fermions,
respectively. 
The term $m_0$ denotes the quark mass appearing in the
relativistic lattice action.
Then the problem appears as a violation of the relation
$m_0=m_1=m_2$ of order $(am_Q)^2$.

One way to avoid the problem of this large $am_Q$ error is
to reinterpret the lattice action using HQET as done in the
Fermilab formulation \cite{El-Khadra:1996mp}.
But, if one does not employ this reinterpretation, one must
restrict oneself to the region where $am_Q$ is small enough
that the remaining errors of order $(am_Q)^2$ and higher are
under control. 
This is possible for the charm quark if one takes the
lattice cutoff much higher than 2~GeV and eliminates the
leading error $(am_Q)^2$ by an extrapolation to the
continuum limit, although it is numerically rather
demanding.
Such analysis has been done only recently for the
charm quark mass \cite{Rolf:2002gu} and for the $D_s$ meson
decay constant \cite{Juttner:2003ns} in the quenched
approximation. 

To reach the bottom quark mass one must rely on an
extrapolation in the heavy quark mass.
It can be done by using the heavy quark scaling law,
\textit{e.g.} for heavy enough quark masses the heavy-light
meson decay constant $f_P$ behaves as $1/\sqrt{M_P}$ up to
power corrections. 
This method requires a careful analysis because the
discretization error of order $(am_Q)^2$ grows as $m_Q$
becomes larger, and this growth could even be amplified by
the extrapolation, producing uncontrolled systematic
errors. 
An extrapolation to the continuum limit $a\to 0$ must be
done \textit{before} the heavy quark extrapolation to reduce 
such a problem.
A combined fit with the result in the static limit may be
used to stabilize the extrapolation.

Because the large energy scale of order $m_Q$ flows only in
the temporal direction in the rest frame, the discretization
error appears from the temporal derivative of the lattice
action.
It is therefore possible to avoid the large discretization
error by taking a temporal lattice spacing $a_t$ much
smaller than $1/m_Q$ while keeping the spatial lattice
spacing $a_s$.
This is called the anisotropic lattice action
\cite{Alford:1996nx}.
Although the radiative correction could reintroduce large
$\alpha_s m_Q a_s$ contributions, in wihch case the virtue
of the anisotropy is lost \cite{Harada:2001ei,Aoki:2001ra},
it is also possible to construct lattice actions free from
this problem \cite{Hashimoto:2003fs}.


\section{Quarkonia}
\label{sec:Quarkonia}

Since the discovery of $J/\psi$, experimental studies of 
heavy quarkonia have produced enormously rich and precise
data for their mass spectra and decay rates.
Theoretically, the quark potential model description has
been successful in explaining their properties, indicating
that heavy quarkonium is a non-relativistic bound
state. 
Therefore, the study of quarkonia gives a firm testing
ground for lattice QCD, and the results may also be used to 
calibrate the basic parameter of the calculation, the
lattice spacing $a$.
Furthermore, the lattice QCD will be useful for studying the 
properties of non-standard states, such as $q\bar{q}g$
hybrids and $q\bar{q}q\bar{q}$ multi-quark states, which may
depend on the details of non-perturbative QCD dynamics.

\subsection{Mass spectrum}
\begin{figure}[tbp]
  \begin{minipage}{8cm}
    \setlength{\unitlength}{.02in}
\begin{picture}(130,150)(5,920)
\thicklines
\put(10,935){\line(0,1){125}}
\multiput(8,950)(0,50){3}{\line(1,0){4}}
\multiput(9,950)(0,10){10}{\line(1,0){2}}
\put(7,950){\makebox(0,0)[r]{9.5}}
\put(7,1000){\makebox(0,0)[r]{10.0}}
\put(7,1050){\makebox(0,0)[r]{10.5}}
\put(7,1065){\makebox(0,0)[r]{GeV}}



\put(27,920){\makebox(0,0)[t]{${^1\rm{}S}_0$}}
\put(25,943.1){\circle*{3}}
\put(30,942){\circle{3}}

\put(52,920){\makebox(0,0)[t]{${^3\rm{}S}_1$}}
\multiput(43,946)(3,0){7}{\line(1,0){2}}
\put(50,946){\circle*{3}}
\put(55,946){\circle{3}}

\multiput(43,1002)(3,0){7}{\line(1,0){2}}
\put(50,1004.1){\circle*{3}}
\put(50,1005.1){\line(0,1){0.2}}
\put(50,1003.1){\line(0,-1){0.2}}
\put(55,1003){\circle{3}}
\put(55,1004){\line(0,1){1.4}}
\put(55,1002){\line(0,-1){1.4}}

\multiput(43,1036)(3,0){7}{\line(1,0){2}}
\put(50,1060){\circle*{3}}
\put(50,1061){\line(0,1){11}}
\put(50,1059){\line(0,-1){11}}
\put(55,1039.1){\circle{3}}
\put(55,1039.1){\line(0,1){7.2}}
\put(55,1039.1){\line(0,-1){7.2}}

\put(92,920){\makebox(0,0)[t]{${^1\rm{}P}_1$}}

\multiput(83,990)(3,0){7}{\line(1,0){2}}
\put(90,987.6){\circle*{3}}
\put(95,989){\circle{3}}
\put(95,990){\line(0,1){0.2}}
\put(95,988){\line(0,-1){0.2}}

\multiput(83,1026)(3,0){7}{\line(1,0){2}}
\put(90,1038.7){\circle*{3}}
\put(90,1039.7){\line(0,1){1.4}}
\put(90,1037.7){\line(0,-1){1.4}}
\put(95,1023){\circle{3}}
\put(95,1023){\line(0,1){7.2}}
\put(95,1023){\line(0,-1){7.2}}

\put(120,920){\makebox(0,0)[t]{${^1\rm{}D}_2$}}
\put(120,1019.2){\circle*{3}}
\put(120,1020.2){\line(0,1){6}}
\put(120,1018.2){\line(0,-1){6}}
\end{picture}

    \vspace*{6mm}
  \end{minipage}
  \begin{minipage}{7.5cm}
    \setlength{\unitlength}{.02in}
\begin{picture}(95,100)(40,-50)

\put(50,-50){\line(0,1){85}}
\multiput(48,-40)(0,20){4}{\line(1,0){4}}
\multiput(49,-40)(0,10){7}{\line(1,0){2}}
\put(47,-40){\makebox(0,0)[r]{$-40$}}
\put(47,-20){\makebox(0,0)[r]{$-20$}}
\put(47,0){\makebox(0,0)[r]{$0$}}
\put(47,20){\makebox(0,0)[r]{$20$}}
\put(47,35){\makebox(0,0)[r]{MeV}}


\put(63,-5){\makebox(0,0)[l]{$h_{\rm b}$}}

\put(70,-0.8){\circle*{3}}
\put(75,-2.9){\circle{3}}
\put(75,-2.9){\line(0,1){1.2}}
\put(75,-2.9){\line(0,-1){1.2}}
\multiput(90,-40)(3,0){7}{\line(1,0){2}}
\put(110,-40){\makebox(0,0)[l]{$\chi_{\rm b0}$}}
\put(97,-24){\circle*{3}}
\put(97,-23){\line(0,1){1}}
\put(97,-25){\line(0,-1){1}}
\put(102,-34){\circle{3}}
\put(102,-34){\line(0,1){5}}
\put(102,-34){\line(0,-1){5}}

\multiput(90,-8)(3,0){7}{\line(1,0){2}}
\put(110,-8){\makebox(0,0)[l]{$\chi_{\rm b1}$}}
\put(97,-8.6){\circle*{3}}
\put(102,-7.9){\circle{3}}
\put(102,-7.9){\line(0,1){2.4}}
\put(102,-7.9){\line(0,-1){2.4}}

\multiput(90,13)(3,0){7}{\line(1,0){2}}
\put(110,13){\makebox(0,0)[l]{$\chi_{\rm b2}$}}
\put(97,10.1){\circle*{3}}

\put(102,11.5){\circle{3}}
\put(102,11.5){\line(0,1){2.4}}
\put(102,11.5){\line(0,-1){2.4}}
\end{picture}

  \end{minipage}
  \caption{
    Bottomonium spectrum from Reference~\cite{Davies:1997mg}. 
    The spin dependent splittings of the 1P states are
    shown separately in the right panel.
    Dashed lines indicate the experimental value. 
    The lattice data are for $N_f=0$ 
    (\textit{filled circles}) and $N_f=2$ 
    (\textit{open circles}).
  }
  \label{fig:Upsilon_spectrum}
\end{figure}
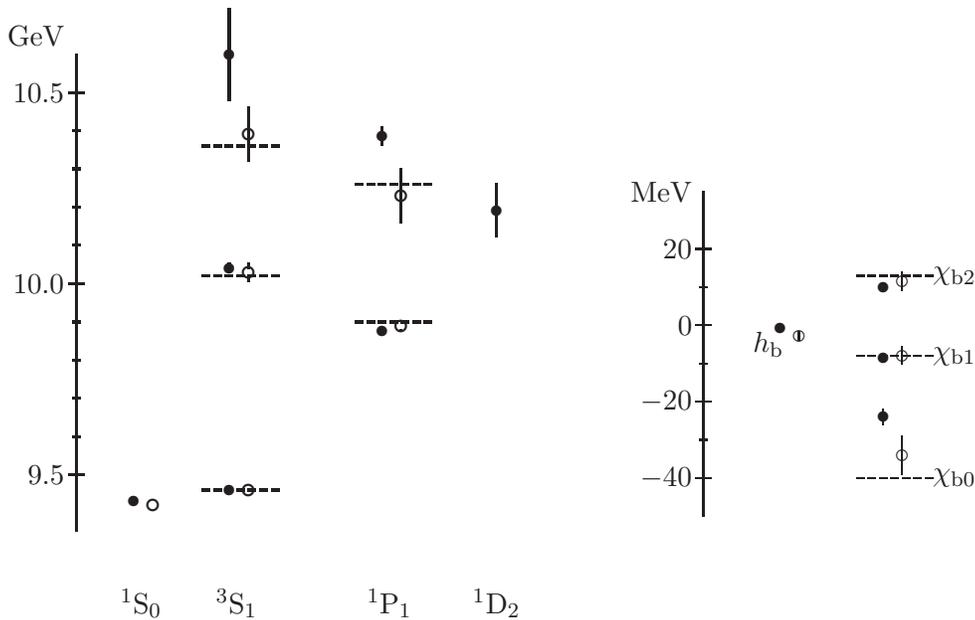

Heavy quarkonium is a system for which one of the most
accurate calculations can be done on the lattice. 
Indeed, Davies \textit{et al.} demonstrated that the
mass spectrum can be precisely calculated using lattice 
NRQCD 
\cite{Davies:1994mp,Davies:1995db,Davies:1997mg,Davies:1998im}.
They used the lattice NRQCD action including the entire
$O(v^4)$ terms as given in Equations~\ref{eq:H_0} and
\ref{eq:delta_H}.
The leading discretization errors are eliminated by adding 
the $O(a^2)$ improvement terms.
The bottomonium spectrum obtained on the lattice with
$1/a\simeq$ 2.4~GeV is shown in
Figure~\ref{fig:Upsilon_spectrum} 
\cite{Davies:1997mg,Davies:1998im}.
The masses of low-lying radial and orbitally excited
states are precisely computed in both quenched ($N_f=0$) and 
unquenched ($N_f=2$) QCD.
It is remarkable that the 1S-2S and 1S-1P splittings
become more consistent with the experimental value if 
dynamical quarks are introduced.

With the NRQCD formulation one can investigate the effect of
each term included in the action by comparing the results
with and without that contribution.
Furthermore, because the bottomonium system is mostly
sensitive to short distance physics, our experience from
quark models and perturbation theory can be used to estimate
the systematic uncertainty due to truncated higher order
terms in $v^2$ and in $a$.

In the non-relativistic expansion the leading terms that 
are not included in the action employed by Davies et al. are
those of $O(v^6)$.  
Because the typical $v^2$ is about 0.1 for the bottomonium
system, the relative error to the leading terms ($\sim v^2$)
is naively $O(v^4)\sim$~1\%. 
An estimate using the quark potential model suggests even
smaller error \cite{Lepage:1991ui}.
The leading discretization errors are the radiative
correction to the $O(a^2)$ improvement term and the $O(a^4)$
corrections.
These are $O(\alpha_s(\pi/a) a^2 p^2)$ and $O(a^4 p^4)$,
respectively, for typical spatial momentum $p=m_Qv$.
For the lattice spacings between 0.05~fm and 0.15~fm the 
size of their corrections is estimated to be at most a few
percent, and lattice data at three lattice spacings support 
such small discretization effects \cite{Davies:1998im}.

For the spin-dependent splittings, the relative error is an
order of magnitude larger because the leading contribution
itself is $O(v^4)$. 
Indeed, Manke \textit{et al.} studied the effect of the
tree-level $O(v^6)$ corrections in quenched NRQCD and found
that the spin-dependent splittings are affected by
10\%--20\% when $O(v^6)$ terms are added to the action,
whereas the spin-averaged splittings are essentially
unaltered,
supporting the above naive expectation \cite{Manke:1997gt}.

As shown in Figure~\ref{fig:Upsilon_spectrum} the mass
spectrum significantly deviates from experiment unless the
effect of dynamical quarks is included.
The sea quark effects have also been studied 
\cite{Eicker:1998vx,Manke:2000dg} and found to be sizable.
The recent calculation by the HPQCD and UKQCD collaborations
on the gauge ensembles with 2+1 flavor sea quarks produced
by the MILC collaboration indicates that both the
spin-averaged splittings and the hyperfine 
splittings are in excellent agreement with experiment
\cite{Davies:2003ik,Davies:2003fw}.

Because the bottomonium spectrum can be precisely calculated 
with good control of the systematic uncertainty, it provides
a reliable input for the lattice scale $1/a$ of the given
lattice. 
Then, other physical quantities can be calculated in the
unit of the bottomonium mass splitting, such as the 1S-1P
splitting.
Among other quantities, the relation between $\alpha_s$ and
the lattice scale $1/a$ can be obtained rather accurately.
Using two-loop perturbation theory to relate $\alpha_s$ on
the lattice with that in the continuum renormalization
scheme, e.g., the $\overline{\mathrm{MS}}$ scheme,
one can determine
$\alpha_{\overline{\mathrm{MS}}}^{(5)}(M_Z)$ using the low
energy hadronic scale as input.
In a recent simulation with 2+1 flavors of dynamical
quarks Davies \textit{et al.} obtained 
$\alpha_{\overline{\mathrm{MS}}}^{(5)}(M_Z)=0.121(3)$
\cite{Davies:2003ik}, which agrees with other experimental
determinations \cite{Hagiwara:fs}.

The charmonium spectrum is much harder to calculate precisely
using the NRQCD formalism, because of larger relativistic
corrections ($v^2\sim$~0.3).
With the same NRQCD action used by Davies \textit{et al.}
the higher order contribution is expected to be about 10\%
for spin-averaged and 30\% for spin-dependent splittings.
In fact the next-to-leading order $O(v^6)$ terms produce a 
sizable effect on the spectrum \cite{Trottier:1996ce}, and
the prescription to (approximately) incorporate radiative
corrections strongly affects the hyperfine splitting
\cite{Shakespeare:1998dt}.
Therefore, relativistic lattice actions have been used
for the charm quark in many studies on anisotropic 
\cite{Chen:2000ej,Okamoto:2001jb} 
and isotropic \cite{Choe:2003wx} lattices.
Extrapolation toward the continuum limit is essential to
control the $O[(am_Q)^2]$ errors in such calculations, and
the unquenched simulation remains to be intractable.

\subsection{Matrix elements}
After successful reproduction of the quarkonium spectra, 
the next interesting problem would be to compute the
quarkonium decay matrix elements in lattice QCD. 
Bodwin et al.~\cite{Bodwin:1994jh} have shown
that the quarkonium decay amplitude can be factorized into 
a product of long-distance matrix elements of four-fermion
operators and short-distance coefficients that are
calculable in perturbation theory.
For instance, the S- and P-wave decay rates are written
as 
\begin{eqnarray}
  \Gamma({}^{2s+1}S_J \rightarrow X) 
  &=&
  \mathcal{G}_1({}^{2s+1}S_J)\, 
  2\,\mathrm{Im} f_1({}^{2s+1}S_J)/m_b^2
  \nonumber\\
  &+ & 
  \mathcal{F}_1({}^{2s+1}S_J)\,
  2\,\mathrm{Im} g_1({}^{2s+1}S_J)/m_b^4,
  \\
  \Gamma({}^{2s+1}P_J \rightarrow X) 
  &=&
  \mathcal{H}_1({}^{2s+1}P_J)\,
  2\,\mathrm{Im} f_1({}^{2s+1}P_J)/m_b^4
  \nonumber\\
  &+&
  \mathcal{H}_8({}^{2s+1}P_J)\,
  2\,\mathrm{Im} g_1({}^{2s+1}P_J)/m_b^4,
\end{eqnarray}
where $f$ and $g$ are the short-distance coefficients 
and the quarkonium states are indicated by ${}^{2s+1}L_J$ 
with spin $s$, orbital angular momentum $L$, and total
angular momentum $J$.
The matrix elements are defined in terms of the bottom quark
field $\psi$ and anti-quark field $\chi$ in the NRQCD
formalism as 
\begin{eqnarray}
  \mathcal{G}_1 
  &=&
  \langle {}^1 S_0 | \psi^{\dagger}\chi \chi^{\dagger}\psi 
  | {}^1 S_0 \rangle,
  \\
  \mathcal{F}_1 
  &=&
  \langle {}^1 S_0 | \psi^{\dagger}\chi \chi^{\dagger}
  (-\frac{i}{2} \overleftrightarrow{\boldvec{D}})^2 \psi 
  | {}^1 S_0 \rangle, 
  \\
  \mathcal{H}_1 
  &=&
  \langle {}^1 P_1 | 
  \psi^{\dagger}\frac{i}{2}\overleftrightarrow{\boldvec{D}}\chi
  \chi^{\dagger}\frac{i}{2}\overleftrightarrow{\boldvec{D}}\psi 
  | {}^1 P_1 \rangle, \\
  \mathcal{H}_8 
  &=&
  \langle {}^1 S_0 | \psi^{\dagger} T^a \chi 
  \chi^{\dagger} T^a \psi   | {}^1 S_0 \rangle,
\end{eqnarray}
where the derivative $\overleftrightarrow{\boldvec{D}}$ is
defined as
$\chi^{\dagger}\overleftrightarrow{\boldvec{D}}\psi=
\chi^{\dagger}(\boldvec{D}\psi)-
(\boldvec{D}\chi)^{\dagger}\psi$.

The lattice calculation of these matrix elements has been
carried out using the $O(v^2)$ lattice NRQCD action for both
quenched \cite{Bodwin:1996tg} and unquenched
\cite{Bodwin:2001mk} QCD.
In order to obtain the matrix element in the continuum NRQCD, 
the lattice operators are matched to those in the continuum 
at the one-loop order of perturbation theory.
Then, the results of $\mathcal{G}_1$ can be compared to the 
phenomenological estimate obtained from the leptonic
decay width 
\begin{equation}
  \Gamma(\Upsilon \rightarrow e^+ e^-)
  \approx \frac{2 \pi Q_b^2 \alpha^2}{3 m_b^2}
  \left(1 -\frac{16 \alpha_s}{3\pi}\right) \mathcal{G}_1
  = 5.0\pm 0.2\;\mathrm{GeV}.
\end{equation}
Bodwin et al. found that the matrix element $\mathcal{G}_1$
in quenched QCD is 40\% smaller than the phenomenological
estimate, 
whereas the unquenched result is consistent with the
phenomenological value within the error of 10\%.
Although further improvements are required, it is
encouraging that one finds good agreement at the present
accuracy of $O(v^2)\sim$ 10 \% level.

\subsection{Exotics}
Most of the known excited states of $b\bar{b}$ and
$c\bar{c}$ mesons are classified in terms of the quark
model; that is, their $J^{PC}$ is explained by the
standard assignment $P=(-1)^{L+1}$ and $C=(-1)^{L+S}$ where 
$S=0$, $1$ is the total spin of $q$ and $\bar{q}$ and $L=0$,
$1$, $2$, ... is the orbital angular momentum.
In QCD, however, there could be other excited states, which
may be explained by a gluon excitation $q\bar{q}g$ or a
four-quark state $q\bar{q}q\bar{q}$.
An experimental candidate of the latter, called $X(3872)$,
was recently found in the charmonium system by the BELLE 
collaboration \cite{Choi:2003ue}.

\begin{figure}[tbp]
  \centering
  \includegraphics[width=8cm]{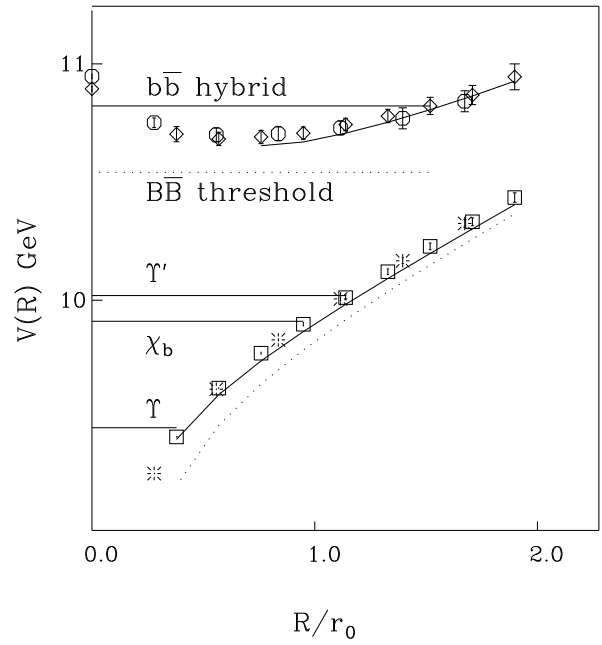}
  \caption{
    Heavy quark potential including gluon excitations.
    The plot is from \cite{Perantonis:1990dy,Michael:2003xg}.
  }
  \label{fig:potential}
\end{figure}

The quark-anti-quark potential has been calculated on the
lattice, including the gluon excitations 
\cite{Perantonis:1990dy,Juge:2002br,Bali:2003jq}.
Figure~\ref{fig:potential} shows such a potential with and 
without first gluonic excitation in the quenched
approximation.
Starting from the measured potential one can calculate the
levels of hybrid mesons by employing some assumptions, such
as the Born-Oppenheimer approximation \cite{Juge:1999ie}.

\begin{figure}[tbp]
  \centering
  \includegraphics*[width=8cm]{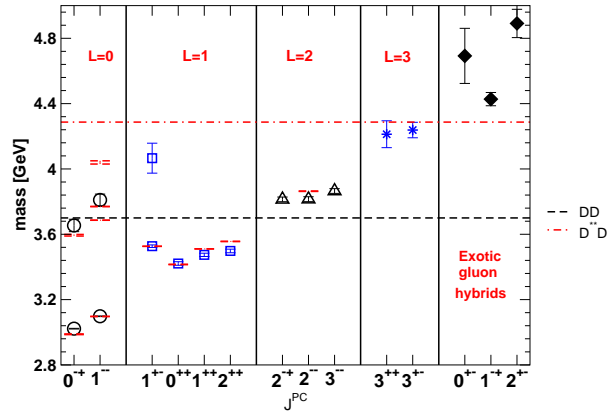}
  \caption{
    Charmonium spectrum from Reference~\cite{Liao:2002rj}.
    The rightmost three points correspond to the hybrid
    states. 
  }
  \label{fig:cchybrids}
\end{figure}

It is also possible to directly calculate the hybrid
spectrum in lattice QCD by choosing interpolating operators
with appropriate quantum numbers.
To obtain a good statistical signal in the lattice
calculation, the anisotropic lattice is useful, and 
such a lattice calculation has been done
\cite{Juge:1999ie,Manke:1998qc,Manke:2001ft,%
  Liao:2001yh,Liao:2002rj}.
The $c\bar{c}$ spectrum obtained by \cite{Liao:2002rj} is
shown in Figure~\ref{fig:cchybrids}.

\section{$B$ physics}
\label{sec:B_physics}

The strongest motivation for the lattice simulation of 
heavy quarks comes from $B$ physics.
One of the goals of the $B$ factory experiments is to
precisely determine the CKM matrix elements, which are
fundamental parameters of the Standard Model.
In particular, the two smallest elements $V_{ub}$ and
$V_{td}$ are interesting because they describe the mixing
between first and third generations and are the source of
$CP$ violation. 
Furthermore, by studying a variety of
flavor-changing-neutral-current (FCNC) processes, which are
suppressed in the Standard Model by the GIM mechanism, one
may get insight into the physics beyond the Standard Model
entering into the loop diagrams.

The bottom quark decay inside the $B$ meson is always
accompanied by a gluon and quark cloud, which makes the
extraction of fundamental parameters from experimental data 
difficult. 
Therefore, a model independent calculation of QCD effects
by a theoretical method, such as lattice QCD, is essential. 

Owing to the practical limitation of computer power, the
momentum scale that lattice QCD can treat is limited to
$\lesssim$ 1~GeV, which is why the effective theories as
discussed in Section~\ref{sec:HQET_and_NRQCD} are developed.
Computer power also limits the momentum of initial and final
state hadrons to be simulated on the lattice to $\lesssim$
1~GeV, whereas in realistic $B$ decays the momentum of
the daughter particles can be as large as $\sim m_B/2$.
This means that only a limited range of kinematical space is 
covered by lattice QCD.
Another important limitation of lattice calculation is that
one cannot treat the multi-hadron final state on the lattice,
because the relation between the correlator in the Euclidean
space and the amplitude in the Minkowski space-time is
non-trivial when more than one particles are involved.
Therefore, the direct calculation of the decay amplitude of 
$B\rightarrow\pi\pi$, for instance, is far beyond the reach
of present lattice calculation. 

In the following section, we review the lattice calculation
of several important $B$ meson matrix elements.

\subsection{$B^0$-$\bar{B}^0$ Mixing}
The mass difference (or oscillation frequency) between two
neutral $B$ mesons $\Delta M_{B_q}$ ($q=d$ or $s$) is given
by 
\begin{equation}
  \Delta M_{B_q}= \frac{G_F^2}{6\pi^2}
  \eta_B M_{B_q} f_{B_q}^2\hat{B}_{B_q} m_W^2 
  S_0(m_t^2/m_W^2) |V_{tq} V_{tb}|^2,
\end{equation}
where $S_0(x)$ is the Inami-Lim function \cite{Inami:1980fz}
originating from the box diagram
and $\eta_B$ is a short distance QCD correction calculated
in perturbative QCD.
The non-perturbative quantity to be computed on the lattice
is the combination, $f_{B_q}^2\hat{B}_{B_q}$.
The leptonic decay constant $f_{B_q}$ is defined through 
\begin{equation}
  i f_{B_q} p_{\mu} = \langle 0 | A_\mu | B_q(p) \rangle,
\end{equation}
with a heavy-light axial-vector current 
$A_\mu=\bar{q} \gamma_5\gamma_\mu b$.
The $B$ parameter $B_{B_q}$ is defined as
\begin{equation}
  \langle \bar{B}^0_q | O^{\Delta B =2}(\mu) | B_q^0 \rangle
  \equiv  \frac{8}{3} B_{B_q}(\mu) f_{B_q}^2 M_{B_q}^2
\end{equation}
with the $\Delta B=2$ operator
\begin{equation}
  O^{\Delta B=2} = 
  \bar{q} \gamma_\mu(1-\gamma_5) b\,
  \bar{q} \gamma_\mu(1-\gamma_5) b.
\end{equation}
This operator has an anomalous dimension, and as a result
the $B$ parameter is scale dependent, $B_{B_q}(\mu)$.
For convenience a scale-invariant $B$ parameter,  
$\hat{B}_{B_q}=B_{B_q} \alpha_s(\mu)^{-6/23}
(1+J_5 \alpha_s/4\pi)$, with $J_5=5165/3174$, is often
quoted.

Experimentally, the mass difference $\Delta M_B$ is already
known very precisely for the $B_d^0$ meson
($\Delta M_B=0.502\pm0.007$~ps$^{-1}$ \cite{HFAG}),
and the main problem in extracting the CKM element $|V_{td}|$
is in the theoretical calculation of $f_B^2 B_B$.
For the $B_s$ meson, to date only a lower bound 
$\Delta M_s >$ 14.4~ps$^{-1}$ (95\% CL) \cite{HFAG} has been
obtained.
The leptonic decay $B^+\to l\nu_l$ is hard to measure at
existing experimental facilities because of its small
branching fraction, although there is a good chance that it
will be measured and thus a combination $|V_{ub}|f_B$ will
be determined at future higher luminosity $B$ factories.

Because the $B_d$ and $B_s$ mesons differ only in the
valence light quark mass as far as QCD is concerned, one can
expect that the theoretical uncertainty largely cancels in
the ratio
\begin{equation}
  \frac{\Delta M_{B_s}}{\Delta M_{B_d}} =
  \frac{|V_{ts}|^2}{|V_{td}|^2}
  \xi^2,\;\;\;\;
  \xi\equiv
  \frac{f_{B_s}\sqrt{B_{B_s}}}{f_{B_d}\sqrt{B_{B_d}}}
\end{equation}
up to the SU(3) breaking effect.
Once the $B_s$-$\bar{B}_s$ mixing rate is experimentally
measured with good precision, it will help to determine the
most interesting CKM matrix element $|V_{td}|$ (assuming the 
CKM unitarity relation $|V_{ts}|=|V_{cb}|$).

The width difference $\Delta\Gamma_s$ of $B_s$ mesons is
induced by the final states into which both $B_s$ and
$\bar{B}_s$ mesons can decay.
(The width difference of $B_d$ in the Standard Model 
is too small to be observed.)
The main contribution comes from 
$\bar{b}s\to c\bar{c} \to b\bar{s}$ transitions at the quark
level.
Because large momentum of order $m_b/2$ flows into final
state particles, the amplitude can be expressed in terms of
the heavy quark expansion \cite{Beneke:1996gn,Beneke:1998sy}
and the leading order is represented by matrix elements of
local $\Delta B=2$ four-quark operators
$O_L=\bar{s} \gamma_\mu(1-\gamma_5) b\,
     \bar{s} \gamma_\mu(1-\gamma_5) b$ and
$O_S=\bar{s}(1-\gamma_5) b\,
     \bar{s}(1-\gamma_5) b$.
The corresponding matrix elements are $B_B$ and $B_S$,
defined such that they are unity in the vacuum saturation
approximation.

\subsection{$B^0$-$\bar{B}^0$ Mixing: Quenched Lattice Results}
In the quenched approximation, the $B$ meson decay constant
has been studied extensively. 
Recent calculations using the Fermilab action
\cite{Aoki:ji,El-Khadra:1997hq,Bernard:1998xi},
the NRQCD action 
\cite{AliKhan:1998df,Ishikawa:1999xu,Collins:2000ix,AliKhan:2001jg}, 
and the extrapolation method 
\cite{Becirevic:1998ua,Bowler:2000xw,Lellouch:2000tw}
show good agreement within an error of $O(15\%)$.
In the recent lattice conferences, world averages read
$f_B=173\pm 23$~MeV, $f_{B_s}=200\pm 20$~MeV, and
$f_{B_s}/f_{B_d}=1.15\pm0.03$
\cite{Ryan:2001ej,Yamada:2002wh}. 
The statistical error of the Monte Carlo simulation is a
minor part of the error given above.
The source of systematic error differs significantly
according to the method used to treat the heavy quark.
In the Fermilab and NRQCD actions, the truncation of the
perturbative and non-relativistic expansions is a major
source of uncertainty.
For the extrapolation method, the matching factor is known
non-perturbatively and the non-relativistic expansion is not
involved, but the extrapolation in the inverse heavy quark
mass is a serious source of systematic error, as we
discussed in Section~\ref{sec:HQET_and_NRQCD}.
In addition, there is an error associated with the scale
setting of the lattice, which is done by using some
reference quantity---such as the rho meson mass, the pion
(or kaon) decay constant, the string tension (or the 
``Sommer scale'' $r_0$ which is also related to the heavy
quark potential \cite{Sommer:1993ce}), and the quarkonium
mass spectra.
In the quenched approximation the lattice scales determined
from different quantities do not necessarily agree. 
This is a potential source of systematic uncertainty.

More recently, there has been an attempt to control the
$O[(am_Q)^2]$ error by performing the simulations on a small  
physical volume, such as (0.4~fm)$^3$, and matching the
results to larger lattices recursively
\cite{deDivitiis:2003wy}. 
The result $f_{B_s}=192\pm 6\pm 4$~MeV agrees with
the previous calculations and the quoted error is
significantly smaller than the present world average.
Another attempt is a further refinement of the extrapolation 
method \cite{Rolf:2003mn}.
By taking the continuum limit at each heavy quark mass,  
Rolf {et al.} eliminate the leading $O[(am_Q)^2]$ error. 
The combined fit with the static result, which is also
extrapolated to the continuum limit, yields a preliminary
result $f_{B_s}=206\pm 10$~MeV \cite{Rolf:2003mn}.
By continuing in these and other directions, it will be
possible to reduce the error in $f_B$ to the 5\% level.

The $B$ parameters have been calculated on quenched
lattices via the static action
\cite{Ewing:1995ih,Christensen:1996sj,Gimenez:1996sk,Gimenez:1998mw},
the NRQCD action
\cite{Hashimoto:1999ck,Hashimoto:2000eh,Aoki:2002bh},
and the extrapolation method
\cite{Becirevic:2000nv,Lellouch:2000tw}, for which the
systematic error is better controlled by a combined fit with
the static results \cite{Becirevic:2001xt}.
Their results are consistent with each other, and the
current average is $\hat{B}_B=1.33\pm 0.12$
\cite{Yamada:2002wh}. 

The other $B$ parameter $B_S$, which is relevant to the
$B_s$ meson width difference, has also been calculated in
the quenched approximation
\cite{Hashimoto:1999yh,Hashimoto:2000eh,Aoki:2002bh,%
  Becirevic:2000sj,Gimenez:2000jj,Flynn:2000hx,%
  Becirevic:2001xt}.
The physical result for $\Delta\Gamma_s/\Gamma_s$ including
next-to-leading order QCD corrections is
$0.074\pm 0.024$ \cite{Ciuchini:2003ww}, which can be
compared to the current experimental average
$0.07^{+0.09}_{-0.07}$ \cite{HFAG}.

\subsection{$B^0$-$\bar{B}^0$ Mixing: Unquenching}
\label{sec:unquenched_BB}
In the past few years, the unquenched calculation of $f_B$
has been performed by several groups 
\cite{Bernard:1998xi,Collins:1999ff,AliKhan:2000eg,AliKhan:2001jg},
and the results seemed larger than the quenched results by
$\sim$ 10\%--15\%.
But the conclusion is now less clear because of an
uncertainty in the chiral extrapolation as we discuss below.

With the present algorithms to incorporate the effect of
dynamical fermions the simulation cost increases as
$1/m_q^3$, where $m_q$ is the light quark mass
\cite{Ukawa:pc}.  
For example, for Wilson-type fermions the smallest
possible sea quark mass is about a half of the strange quark 
mass $m_s$ even on the fastest available supercomputer. 
Hence, one must extrapolate the lattice data from the light
quark mass region $m_q\gtrsim m_s/2$ toward the physical up
and down quark masses, which are about $m_s/25$. 
In such an extrapolation the theoretical guide to restrict
its functional form is given by chiral perturbation
theory (ChPT) \cite{Gasser:1983yg,Gasser:1984gg}, which
provides a systematic framework to calculate physical
quantities as an expansion around the massless limit with an 
expansion parameter $m_\pi^2\propto m_q$.
At next-to-leading (one-loop) order, ChPT predicts 
non-analytic behavior coming from pion loops.
For instance, for the pion decay constant the non-analytic
term is
$-\frac{N_f}{2}\frac{m_\pi^2}{(4\pi f)^2}
\ln\frac{m_\pi^2}{\mu^2}$
which is called the chiral logarithm.
($f$ is the pion decay constant in the massless limit.)
Its coefficient depends only on the number of dynamically
active quark flavors $N_f$, which is a consequence
of the chiral symmetry of QCD.

The chiral logarithm could complicate the analysis of
lattice data, because the effect of such non-analytic
functional dependence can be missed if one simply relies on
a polynomial fit of lattice data. 
Furthermore, when the lattice data are available only in the
relatively heavy mass region $m_q\gtrsim m_s/2$, the
convergence of the chiral expansion itself is questionable. 
An explicit study using $N_f=2$ lattice data from
Reference~\cite{Aoki:2002uc} suggests that the region 
$m_q\gtrsim m_s/2$ is beyond the reach of ChPT 
\cite{Hashimoto:2002vi}.
If this is true, one must introduce some model function to
fit the lattice data, in order to make the chiral
extrapolation consistent with ChPT, which introduces
significant uncertainty in the chiral extrapolation. 
\footnote{
  This point requires further investigation.
  Because the Wilson-type fermions explicitly violate the
  chiral symmetry, it is not clear a priori whether the lattice
  data can be fitted with the continuum ChPT formula.
  An extension of ChPT that includes such explicit chiral
  symmetry breaking has been proposed for the Wilson-type
  fermions 
  \cite{Rupak:2002sm,Bar:2003mh,Aoki:2003yv} and for the
  staggered fermions
  \cite{Lee:1999zx,Aubin:2003mg,Aubin:2003uc}.
}

The heavy mesons can be incorporated into the framework of
ChPT. 
At next-to-leading order \cite{Grinstein:1992qt},
the $B$ meson decay constant depends on the light quark mass
as follows:
\begin{equation}
  \frac{\Phi_{f_B}}{\Phi_{f_B}^{(0)}} = 
  1 - \frac{3(1+3g^2)}{4} 
  \frac{m_\pi^2}{(4\pi f)^2} \ln \frac{m_\pi^2}{\mu^2}
  + c_1 m_\pi^2 + \cdots.
  \label{eq:ChPT_one-loop_fB}
\end{equation}
Here, $\Phi_{f_B}\equiv f_B\sqrt{M_B}$ and
$\Phi_{f_B}^{(0)}$ denotes its value in the massless light
quark limit.
The coupling $g$ describes the $B^*B\pi$ interaction in
ChPT, which is experimentally measured through the 
$D^*\to D\pi$ decay as $g=0.59\pm 0.01\pm 0.07$
\cite{Ahmed:2001xc,Anastassov:2001cw}, if heavy quark
symmetry is assumed.

\begin{figure}[tbp]
  \centering
  \includegraphics*[width=10cm,clip=true]{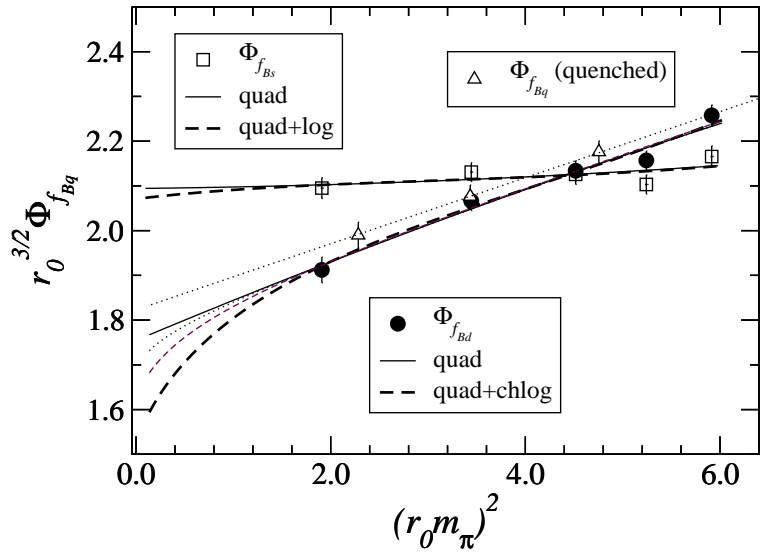}
  \caption{
    Chiral extrapolation of $\Phi_{f_{B_d}}$ 
    (\textit{filled circles}) 
    and $\Phi_{f_{B_s}}$ (\textit{open squares}).
    A polynomial fit is shown by solid lines; 
    the fits with the hard cutoff chiral logarithm are
    shown for $\mu$ = 300 (\textit{dotted curve}), 500
    (\textit{thin dashed curve}) and $\infty$ 
    (\textit{thick dashed curve}) MeV. 
    Quenched results are also shown (\textit{triangles}).
    The plot is from Reference~\cite{Aoki:2003xb}.
  }
  \label{fig:chi_fB}
\end{figure}

Figure~\ref{fig:chi_fB} shows the chiral extrapolation of
$\Phi_{f_{B_d}}$. 
The lattice data are from a recent unquenched simulation
with two flavors of the $O(a)$-improved Wilson fermion
\cite{Aoki:2003xb}. 
The NRQCD action is used for the heavy quark.
In order to estimate the systematic uncertainty associated
with the chiral logarithm, the lattice results are fitted
with Equation~\ref{eq:ChPT_one-loop_fB}, but the chiral
logarithm term is modified as 
$m_\pi^2\ln m_\pi^2/\mu^2 \rightarrow
 m_\pi^2\ln m_\pi^2/(m_\pi^2+\mu^2)$.
This functional form is motivated by a hard cutoff
regularization of one-loop ChPT;
it is designed such that the effect of the pion loop is
suppressed above the cutoff scale $\mu$, whereas 
Equation~\ref{eq:ChPT_one-loop_fB} is reproduced when 
$m_\pi^2\ll \mu^2$.
By taking a maximum parameter range $\mu=0\to\infty$~MeV,
the results obtained are 
$f_{B_d}=191(10)(^{+12}_{-22})$~MeV,
$f_{B_s}=215(9)(^{+14}_{-13})$~MeV, and
$f_{B_s}/f_{B_d}=1.13(3)(^{+13}_{-\ 2})$,
where the first error is statistical, and the second is
the estimate of the uncertainty due to chiral extrapolation
and other sources added in quadrature \cite{Aoki:2003xb}.
Because the $f_{B_s}$ is hardly affected by the chiral
logarithm, a large uncertainty remains only in $f_{B_d}$.
The SU(3) breaking ratio $f_{B_s}/f_{B_d}$ contains
significantly larger error than what was often reported in
the quenched calculation, for which the effect of chiral
logarithm was ignored.

The problem of large uncertainty due to the chiral logarithm
was also emphasized in
References~\cite{Yamada:2001xp,Kronfeld:2002ab}. 
A possible way to reduce the error is to extrapolate a ratio 
$f_B/f_\pi$ instead of the individual decay constants
\cite{Becirevic:2002mh}.
Because the coefficient of the chiral logarithm terms 
in $f_B$ and $f_{\pi}$ are numerically close, the
uncertainty from the extrapolation will become smaller.
For the cancellation of the chiral logarithms, a more
natural choice is to consider $f_B/f_D$ or the Grinstein
ratio $(f_{B_s}/f_B)/(f_{D_s}/f_D)$ \cite{Grinstein:1993ys},
in which the chiral logarithm exactly cancels at the leading
order of the $1/M$ expansion.
These ratios are useful because the CLEO-c, an experimental
progrom at Cornell, promises to measure $f_D$ and $f_{D_s}$
with a precision of a few per cent in a few years. 
An unquenched lattice calculation of the Grinstein ratio has 
already been attempted aiming at a few per cent accuracy
\cite{Onogi:jn}.

To be convincing, however, the unquenched simulations must
employ much smaller sea quark masses.
With currently available computational resources, only
staggered fermions are feasible as dynamical fermions to
carry out such simulations.
The first such calculation including the dynamical effects
of one strange and two light quarks was presented
recently for $f_{B_s}$ \cite{Wingate:2003gm}.
The lightest sea quark mass is as small as $m_s/4$ in this
work.
The result, $f_{B_s}=260\pm 29$~MeV, is somewhat larger than
the previous two-flavor result.

The $B$ parameter is also calculated in
Reference~\cite{Aoki:2003xb}. 
The chiral logarithm is less problematic here, because the 
coefficient of the one-loop chiral log term is 
$-(1-3g^2)/2$ rather than $3(1+3g^2)/4$ in
Equation~\ref{eq:ChPT_one-loop_fB} \cite{Sharpe:1995qp} and
is numerically negligible in practice.
Results for the physically relevant quantities are
$f_{B_d}\sqrt{\hat{B}_{B_d}}=215(11)(^{+15}_{-27})$~MeV,
$f_{B_s}\sqrt{\hat{B}_{B_s}}=245(10)(^{+19}_{-17})$~MeV,
and
$\xi= 1.14(3)(^{+13}_{-\ 2})$.

\subsection{$B\to D^{(*)}l\nu$ Form Factors}
The semi-leptonic decays $b\to cl\nu$ and $b\to ul\nu$ may be
used to determine the CKM matrix elements $|V_{cb}|$ and
$|V_{ub}|$ respectively, if their decay rates are
theoretically calculated in a model independent way.
For the exclusive decay modes, such as $B\to D^{(*)}l\nu$
and $B\to\pi l\nu$, the quantities to be calculated are
their form factors.

For the heavy-to-heavy decays, namely $B\to D^{(*)}l\nu$,
one can use the heavy quark symmetry to restrict their form 
factors, \textit{i.e.} several form factors with different
spin structures are related to one universal form factor
called the Isgur-Wise function \cite{Isgur:vq,Isgur:ed}.
Furthermore, thanks to the symmetry between the initial and 
final states, the form factor is normalized to unity in the 
kinematical end point where the daughter $D^{(*)}$ meson
does not have spatial momentum in the rest frame of the
initial $B$ meson.
Specifically, if one writes the differential decay rate in
terms of $w\equiv v\cdot v'$ ($v$ and $v'$ are the
four-velocity of $B$ and $D^{*}$ mesons, respectively)
\begin{equation}
  \frac{d\Gamma(B\to D^{*}l\nu)}{dw}
  = \frac{G_F^2 |V_{cb}|^2 m_B^5}{48\pi^3}
  \mathcal{K}(w) \mathcal{F}(w)^2
\end{equation}
with $\mathcal{K}(w)$ a known kinematical factor, then the form
factor $\mathcal{F}(w)$ obeys $\mathcal{F}(1)=1$.
The CKM element $|V_{cb}|$ can thus be determined through
the $w\to 1$ limit of experimental data.

However, because the heavy quark symmetry is exact only in
the limit where both heavy quarks have infinite mass, one
must take the $1/m$ corrections into account.
Again using the heavy quark symmetry one can show that the 
leading correction to the relevant form factor $h_{A_1}(1)$,
which is identical to $\mathcal{F}(1)$ in the zero-recoil 
limit, is $O(1/m^2)$ \cite{Luke:1990eg}:
\begin{equation}
  h_{A_1}(1) = \eta_A [1+\delta_{1/m^2}+\delta_{1/m^3}],
\end{equation}
with $\eta_A$ a short distance correction that comes from
the matching between QCD and HQET, and
\begin{equation}
  \delta_{1/m^2} =
  - \frac{\ell_V}{(2m_c)^2}  + \frac{2 \ell_A}{2m_c 2m_b}
  - \frac{\ell_P}{(2m_b)^2}.
\end{equation}
The parameters $\ell_P$, $\ell_V$, and $\ell_A$ are
non-perturbative quantities, which were previously
estimated using the non-relativistic quark model
\cite{Falk:1992wt,Neubert:1994vy}.

These three parameters appear in the $1/m$ expansions of
other form factors because of the heavy quark symmetry
\cite{Falk:1992wt,Mannel:1994kv},
and thus can be obtained from lattice calculation of
double ratios of zero recoil matrix elements
\cite{Hashimoto:2001nb,Kronfeld:2000ck}
\begin{eqnarray}
  \frac{
    \langle D|\bar{c}\gamma^4 b|\bar{B}\rangle
    \langle\bar{B}|\bar{b}\gamma^4 c|D\rangle}{
    \langle D|\bar{c}\gamma^4 c|D\rangle
    \langle\bar{B}|\bar{b}\gamma^4 b|\bar{B}\rangle
  }
  & = &
  \left\{ \eta_V^{\mathrm{lat}}
    [ 1- \ell_P \Delta + O(1/m^3) ]
  \right\}^2,
  \\
  \frac{
    \langle D^*|\bar{c}\gamma^4 b|\bar{B}^*\rangle
    \langle\bar{B}^*|\bar{b}\gamma^4 c|D^*\rangle}{
    \langle D^*|\bar{c}\gamma^4 c|D^*\rangle
    \langle\bar{B}^*|\bar{b}\gamma^4 b|\bar{B}^*\rangle
  },
  & = &
  \left\{ \eta_V^{\mathrm{lat}}
    [ 1- \ell_V \Delta + O(1/m^3) ]
  \right\}^2,
  \\
  \frac{
    \langle D^*|\bar{c}\gamma^j\gamma^5 b|\bar{B}\rangle
    \langle\bar{B}^*|\bar{b}\gamma^j\gamma^5 c|D\rangle}{
    \langle D^*|\bar{c}\gamma^j\gamma^5 c|D\rangle
    \langle\bar{B}^*|\bar{b}\gamma^j\gamma^5 b|\bar{B}\rangle
  }
  & = &
  \left\{ \eta_A^{\mathrm{lat}}
    [ 1- \ell_A \Delta + O(1/m^3) ]
  \right\}^2,
\end{eqnarray}
where $\Delta\equiv (1/(2m_c)-1/(2m_b))$.
The terms $\eta_V^{\mathrm{lat}}$ and
$\eta_A^{\mathrm{lat}}$ are matching factors between lattice
theory and continuum HQET \cite{Harada:2001fj}. 
The double ratios of matrix elements can be precisely
calculated, because both statistical and systematic errors
largely cancel in the ratio.
In the quenched approximation,
$\mathcal{F}(1)=0.913^{+0.024+0.017}_{-0.017-0.030}$ 
has been obtained \cite{Hashimoto:2001nb}. 
This may be compared to the quark model result
\cite{Neubert:1994vy} combined with a two-loop calculation
of $\eta_A$ \cite{Czarnecki:1996gu,Czarnecki:1997cf},
namely $0.907\pm 0.031$, or 
the QCD sum rule result
\cite{Shifman:jh,Bigi:ga,Uraltsev:2000qw}
$0.900\pm 0.038$.

The corresponding form factor $\mathcal{G}(1)$ for 
$B\to Dl\nu$ is written as
\begin{equation}
  \mathcal{G}(1) = h_+(1) - \frac{m_B-m_D}{m_B+m_D} h_-(1), 
\end{equation}
using the form factors $h_+$ and $h_-$ defined through
\begin{equation}
  \langle D(v')|\bar{c}\gamma^\mu b|B(v)\rangle
  =\sqrt{m_B m_D}
  [(v+v') h_+(w) + (v-v') h_-(w)].
\end{equation}
The lattice calculation is more involved because it contains 
$h_-(1)$, which cannot be obtained merely from the zero
recoil matrix elements, but is possible using a similar
technique \cite{Hashimoto:1999yp}.

The Isgur-Wise function $\xi(w)$, which is a heavy quark
limit of $h_+(w)$, can also be calculated on the lattice as
a function of $w$.
The lattice version of HQET that includes finite velocity was
developed for this purpose \cite{Mandula:ds} and was
extended to include $1/m$ corrections 
\cite{Hashimoto:1995in,Sloan:1997fc,Foley:2002qv}.
It is technically more difficult than the usual HQET, because 
the velocity receives finite renormalization, which must 
be determined from non-perturbative simulation
\cite{Hashimoto:1995in,Mandula:1997hb,Christensen:1999gv}.
Lattice calculations using the relativistic fermion have
also been attempted \cite{Bernard:1993ey,Bowler:1995bp}, but
the control of systematic error at the level of a few percent 
(the precision necessary for the determination of $|V_{cb}|$) 
is still challenging.

\subsection{$B\to\pi l \nu$ form factors} 
\label{sec:Btopi}
Because the symmetry between the initial and final states is
lost in the heavy-to-light decays, the form factors are no
longer normalized for $B\to\pi l\nu$ or $\rho l\nu$.
Therefore, lattice QCD must deal with the form factors
themselves rather than just their $1/m$ expansion
coefficients.

The $B\to\pi l\nu$ form factors $f^+(q^2)$ and $f^0(q^2)$
are defined through 
\begin{equation}
  \langle\pi(k)|\bar{q}\gamma^{\mu}b|B(p)\rangle =
  f^+(q^2)
  \left[
    (p+k)^\mu - \frac{m_B^2-m_\pi^2}{q^2} q^\mu
  \right] 
  + f^0(q^2) \frac{m_B^2-m_{\pi}^2}{q^2} q^\mu,
\end{equation}
where $p$ and $k$ are the momenta of the initial $B$ and
final $\pi$ mesons respectively, and $q^\mu=(p-k)^\mu$ is a
momentum transfer to the lepton pair.
The minimum value of $q^2$, namely zero, occurs when the
lepton and neutrino are parallel and the pion is
energetically propelled to the opposite direction.
The maximum value $q^2_{\mathrm{max}}=(m_B-m_\pi)^2$ occurs  
when the lepton and neutrino are emitted back-to-back with
maximum energy and the pion stays at rest. 
The differential decay rate is written as
\begin{equation}
  \label{eq:Btopi_differential_decay_rate}
  \frac{d\Gamma(B\to\pi l\nu)}{dq^2} = 
  \frac{G_F^2}{24\pi^3}
  |V_{ub}|^2  [(v\cdot k)^2-m_\pi^2]^{3/2} 
  |f^+(q^2)|^2.
\end{equation}

Because the discretization error becomes uncontrollable for 
momenta much larger than $\Lambda_{\mathrm{QCD}}$, the
calculation of the form factors is feasible only in the
large $q^2$ (small recoil) region ($q^2\gtrsim$ 15~GeV$^2$),
where the spatial momentum of the pion is lower than roughly 
1~GeV/c. 
The CKM matrix element $|V_{ub}|$ can be obtained by
combining the experimental data integrated above some $q^2$
value, say 15~GeV$^2$, and the lattice results for the form
factor $|f^+(q^2)|^2$ integrated in the same region with an
appropriate kinematical factor.

Extrapolation or interpolation in the heavy quark mass is
done according to the heavy quark scaling law
$f^+(q^2)\sim\sqrt{m_B}$ and $f^0(q^2)\sim 1/\sqrt{m_B}$.
The analysis can be more explicit if one works with
the HQET motivated form factors $f_1(v\cdot k)$ and
$f_2(v\cdot k)$. 
These are defined in the heavy quark limit as
\cite{Burdman:1993es}
\begin{equation}
  \langle\pi(k)|\bar{q}\gamma^\mu b|B(v)\rangle = 2
  \left[
    f_1(v\cdot k) v^\mu + f_2(v\cdot k)\frac{k^\mu}{v\cdot k}
  \right]
\end{equation}
where $v^\mu=p^\mu/m_B$ is the four-velocity of the $B$ meson.
$f^+(q^2)$ and $f^0(q^2)$ are given by a linear combination
of $f_1(v\cdot k)$ and $f_2(v\cdot k)$.

\begin{figure}[tb]
  \centering
  \includegraphics*[width=10cm]{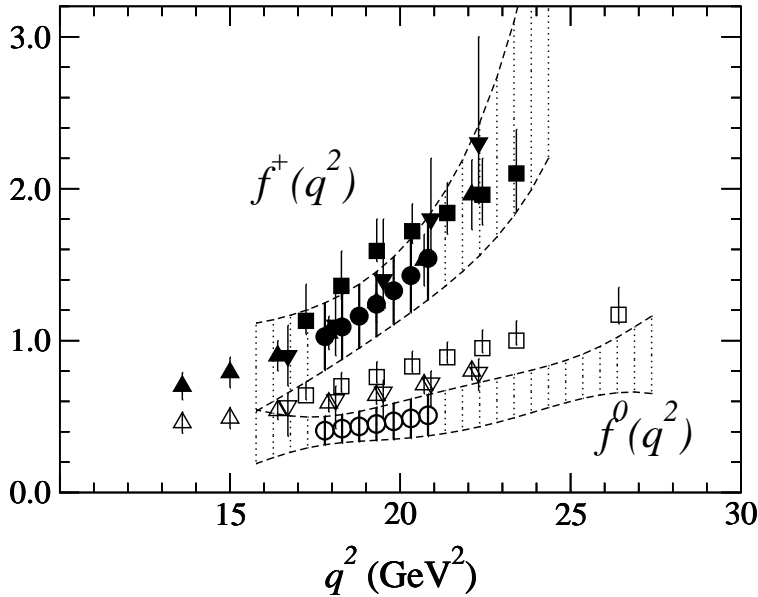}
  \caption{
    $q^2$ dependence of $B\rightarrow\pi l\nu$ form factors
    $f^+(q^2)$ (\textit{filled symbols}) and 
    $f^0(q^2)$ (\textit{open symbols}).
    Upward triangles, downward triangles, squares, circles
    denote data from the 
    UKQCD \cite{Bowler:1999xn}, APE \cite{Abada:2000ty}, 
    Fermilab \cite{El-Khadra:2001rv} and 
    JLQCD \cite{Aoki:2001rd} collaborations, respectively. 
  }
  \label{fig:f+f0}
\end{figure}

Recently, five groups have carried out extensive quenched
calculations of $B\to\pi l\nu$ form factors using the
extrapolation method \cite{Bowler:1999xn,Abada:2000ty},
the Fermilab action \cite{El-Khadra:2001rv}, and the NRQCD
action \cite{Aoki:2001rd,Shigemitsu:2002wh} for heavy
quarks. 
Results for the form factors $f^+(q^2)$ and $f^0(q^2)$ are
shown in Figure~\ref{fig:f+f0}.
Overall, results for $f^+(q^2)$ are in agreement among four
different groups within the error of order 20\%, but some
disagreement is observed for $f^0(q^2)$.
It originates from differences in the light quark mass
dependence of the data. 
Further understanding is necessary to resolve these
differences. 

The chiral extrapolation is more problematic for these form 
factors than for the decay constant, since the reference
point $q^2$ (or $v\cdot k$) also varies with the light quark
mass. 
If one sticks to a particular kinematical point, e.g., the
zero recoil, the extrapolation must include a linear term
in $m_\pi$ in addition to the usual $m_\pi^2$ term.
In fact, the soft pion relation $f^0(q^2)=f_B/f_\pi$ is
poorly satisfied unless the extrapolation includes the
linear $m_\pi$ term \cite{Aoki:2001rd}.
Furthermore, the chiral logarithm
\cite{Fleischer:1992tn,Becirevic:2002sc} must be taken
into account in future unquenched calculations.
Some unquenched studies are in progress, especially using
staggered dynamical quarks
\cite{Shigemitsu:2003xw,Bernard:2003gu,Okamoto:2003ur}.

\subsection{$B\to\rho l \nu$ and other form factors}
The decay $B\to\rho l\nu$ can also be used to determine
$|V_{ub}|$.
However, the lattice calculation of its form factors is
more complicated than for $B\to\pi l\nu$, because 
there are four (not two) independent form factors
corresponding to different spin-momentum combinations. 
Furthermore, the statistical signal of the Monte Carlo
simulation is significantly worse for the rho meson, 
especially when finite spatial momentum is injected.
A more fundamental problem is that the rho meson is an
unstable particle that has a large width $\sim$ 150~MeV.
Lattice simulations of such an unstable particles should
eventually treat the two-pion final states and take its
phase shift into account, the practical feasibility of which
is still an open question. 

\begin{figure}[tb]
  \centering
  \includegraphics[width=10cm]{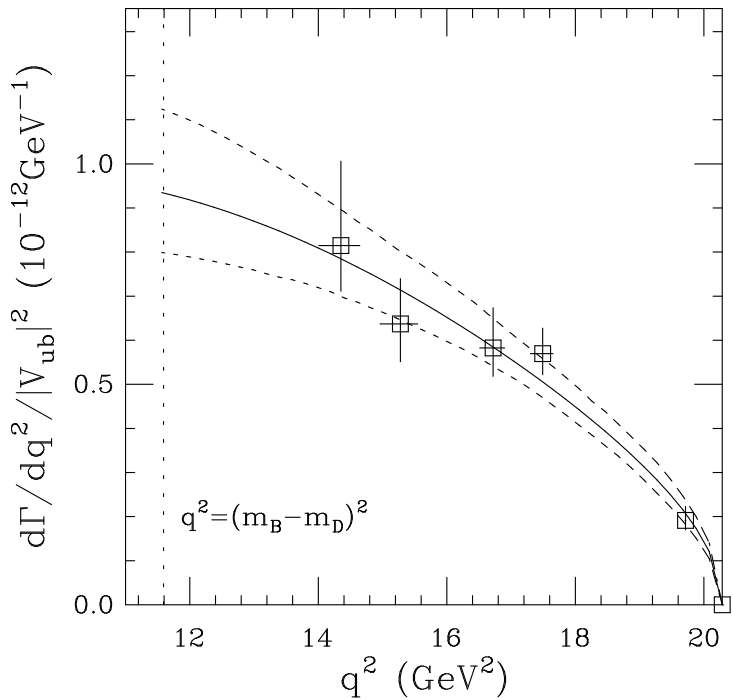}
  \caption{
    Differential decay rate of the $B\to\rho l\nu$ mode.
    Lattice data are shown with a fit curve discussed in the
    text.
    (From Reference~\cite{Flynn:1995dc} with permission).
  }
  \label{fig:Btorho}
\end{figure}

Despite these problems, it is worth calculating those
form factors on the lattice because the calculation provides
a firmer theoretical guide for the experimental analysis
than do the previous model calculations.
A study was made by the UKQCD collaboration using the
extrapolation method \cite{Flynn:1995dc}. 
(Earlier works on $D\to K^{(*)}l\nu$ decays inculde
References
\cite{Lubicz:1991bi,Bernard:fg,Bernard:bz,Bowler:1994zr}.)
As in the $B\to\pi l\nu$ form factors the lattice
calculation is reliable only in the high $q^2$ region.
The extraction of $|V_{ub}|$ can be made by combining a
partially integrated decay rate with the lattice results in 
the corresponding region.
The differential decay rate is written as
\begin{equation}
  \frac{d\Gamma(B\to\rho l\nu)}{dq^2} =
  \frac{G_F^2 |V_{ub}|^2}{192\pi^3 m_B^3}
  q^2 [\lambda(q^2)]^{1/2}
  \left(
    |H^+(q^2)|^2 + |H^-(q^2)|^2 + |H^0(q^2)|^2
  \right),
\end{equation}
where $\lambda(q^2)$ is a kinematical factor.
$H^0(q^2)$ and $H^\pm(q^2)$ denote amplitudes with
longitudinal and transverse rho meson polarization,
respectively, and are written in terms of three form factors 
$A_1(q^2)$, $A_2(q^2)$ and $V(q^2)$.
Near the $q^2_{\mathrm{max}}$ one can parametrize these
amplitudes as $a^2(1+b(q^2-q^2_{\mathrm{max}}))$.
Figure~\ref{fig:Btorho} shows the lattice results for
the differential decay rate with a fit curve using the above 
parametrization.
It demonstrates how lattice calculation can be used to
compare theory with experiments. 

The systematic error is still as large as $\sim$ 25\% from 
the $1/M$ extrapolation and other sources.
The calculation is in the quenched approximation and the
chiral limit of the valence light quark is yet to be taken.
These problems are being addressed.
Indeed, a couple of large scale quenched lattice
calculations are in progress using the extrapolation method 
\cite{Gill:2001jp,Abada:2002ie}, and they have already
reported preliminary results. 

Phenomenologically, the FCNC processes $b\to s(d)\gamma$ and  
$b\to s(d)l^+l^-$ are also interesting because they occur
through penguin and box diagrams, which are sensitive to
possible new-physics contributions.
Their exclusive decay modes are $B\to K^*\gamma$,
$B\to\rho\gamma$, $B\to K^{(*)}l^+l^-$, and 
$B\to K^{(*)}\nu\bar{\nu}$.
For the radiative decays $B\to K^*\gamma$ and $\rho\gamma$
the lattice calculation of the form factor necessarily
involves some model dependence, since the form factor at
$q^2=0$ is relevant.
One must therefore extrapolate the lattice data from the
high $q^2$ region assuming some functional form of the $q^2$ 
dependence, as was done in
References~\cite{Bowler:cq,Burford:1995fc}.

For the $B\to K^{(*)}l^+l^-$ and $B\to K^{(*)}\nu\bar{\nu}$
decays, on the other hand, the lattice calculation may be
useful in the large lepton invariant mass region.
However, the long distance effect due to $q\bar{q}$ and
$c\bar{c}$ resonances must be taken into account for the 
$l^+l^-$ final states.

\subsection{HQET Parameters} 
The HQET parameters, defined as
\begin{eqnarray}
  \mu_{\pi}^2(H_Q) 
  & \equiv &
  \frac{1}{2M_{H_Q}}
  \left\langle H_Q \left| \bar{Q}(i\boldvec{D})^2 Q \right| H_Q \right\rangle,
  \label{eq:mu_pi}
  \\
  \mu_G^2(H_Q) 
  & \equiv &
  \frac{1}{2M_{H_Q}}
  \left\langle H_Q \left| \bar{Q}\boldvec{\sigma}\cdot\boldvec{B} Q \right| H_Q
  \right\rangle,
  \label{eq:mu_G}
\end{eqnarray}
appear in the analysis of the heavy quark
expansion \cite{Neubert:1997gu,Bigi:1997fj}.\footnote{ 
  Another notation, $\lambda_1=-\mu_\pi^2(B)$ and
  $\lambda_2=\mu_G^2(B)/3$, is often used in the literature.
}
For instance, the inclusive decay rate of a heavy hadron
$H_Q$ is written as 
\begin{equation}
  \label{eq:inclusive_decay_rate}
  \Gamma(H_Q\rightarrow X_f) =
  \frac{G_F^2 m_Q^5}{192\pi^3}
  \left[
    c_3^f
    \left(
      1 - \frac{\mu_{\pi}^2(H_Q)-\mu_G^2(H_Q)}{2m_Q^2}
    \right)
    + 2 c_5^f\,
    \frac{\mu_G^2(H_Q)}{m_Q^2}
    + \cdots
  \right],
\end{equation}
where the coefficients $c_3^f$ and $c_5^f$ are
perturbatively calculable
\cite{Chay:1990da,Bigi:1993fe,Manohar:1993qn,Blok:1993va}.
Because the leading order term is independent of the type of 
hadrons, it gives interesting predictions of lifetime
ratios: 
\begin{equation}
  \label{eq:lifetime_ratio}
  \frac{\tau(H_b^{(1)})}{\tau(H_b^{(2)})}
  = 1 + \frac{\mu_\pi^2(H_b^{(1)})-\mu_\pi^2(H_b^{(2)})}{2m_b^2}
  + c_G \frac{\mu_G^2(H_b^{(1)})-\mu_G^2(H_b^{(2)})}{m_b^2}
  + O\left(\frac{1}{m_b^3}\right),
\end{equation}
with a perturbative coefficient $c_G\simeq$ 1.2
\cite{Neubert:1996we}.

The parameters $\mu_{\pi}^2(H_Q)$ and $\mu_G^2(H_Q)$ are
non-perturbative parameters that depend on heavy hadrons
$H_Q$. 
Whereas the spin-dependent parameter $\mu_G^2(H_Q)$ is
reliably determined from the hyperfine splitting
$m_{B^*}-m_B$, the determination of $\mu_\pi^2(H_Q)$
requires non-perturbative methods, such as lattice QCD. 

The lattice calculation of $\mu_\pi^2(H_Q)$ suffers from the 
subtraction of the quadratic divergence that appears in the
matching of the lattice operator $-\bar{Q}\boldvec{D}^2 Q$
to its continuum counterpart.
Because the perturbative subtraction involves large
systematic error \cite{Martinelli:1996pk}, a
non-perturbative subtraction has been attempted and the
result $\lambda_1=0.09\pm 0.14$~GeV$^2$ has been obtained 
\cite{Crisafulli:1995pg,Gimenez:1996av}.
Another possible approach is to rely on the mass formula,
such as $M_{\bar{B}}=m_b+\bar{\Lambda}-\lambda_1/2m_b$ for a
spin-averaged 1S $B$ meson, and to fit the lattice data as a
function of $1/m_b$.
The results of such analysis using the NRQCD action
\cite{AliKhan:1999yb} and the Fermilab action
\cite{Kronfeld:2000gk} are
$\lambda_1=-0.1\pm 0.4$~GeV$^2$ and $-0.45\pm 0.12$~GeV$^2$,
respectively.
In this method the quadratic divergence is subtracted away
by the fitting, but the matching of the kinetic term in
the non-relativistic expansion is a possible source of
error.
The reason for the disagreement with the above result with
non-perturbative subtraction is not clear yet.

It is also possible to avoid the subtraction of the
quadratic divergence by concentrating on the difference
among different hadrons, \textit{e.g.} 
$\mu_\pi^2(\Lambda_b)-\mu_\pi^2(B)$,
which is still useful in the estimate of lifetime ratios
through Equation~\ref{eq:lifetime_ratio}.
A recent quenched calculation \cite{Aoki:2003jf}, 
$\mu_\pi^2(\Lambda_b)-\mu_\pi^2(B)=-0.01\pm 0.52$~GeV$^2$, 
is consistent with most phenomenological estimates.
In particular, the well-known inconsistency of the heavy
quark expansion of $\tau(\Lambda_b)/\tau(B)\simeq 0.98$ with
the experimental value $0.77\pm 0.05$ \cite{HFAG}
continues to be a problem. 
At the $1/m_b^3$ order, the spectator quark effect, which is
expressed in terms of $\Delta B=0$ four-quark operators,
becomes important \cite{Neubert:1996we}.
A lattice calculation \cite{DiPierro:1999tb} of those matrix 
elements suggests that the spectator effects are indeed
significant but not sufficiently large to account for the
full discrepancy.

\subsection{$B^* B\pi$ Coupling}
The heavy quark symmetry and chiral symmetry can be combined 
into the framework of heavy meson ChPT 
\cite{Burdman:gh,Wise:hn,Yan:gz}, which enables us to
systematically expand physical amplitudes in terms of small 
pion momenta. 
This theory includes one additional parameter, $g$. 
It is related to the $B^*B\pi$ coupling $g_{B^*B\pi}$
defined by
\begin{equation}
  \label{eq:B*Bpi}
  \langle B(p)\pi(q) | B^*(p') \rangle 
  = - g_{B^*B\pi} q_\mu \eta^\mu 
  (2\pi)^4 \delta^4(p'-p-q),
\end{equation}
where $\eta^\mu$ is the polarization vector of the $B^*$
meson, and the $B^*B\pi$ coupling $g_{B^*B\pi}$ is
proportional to the low energy constant $g$ as 
$g_{B^*B\pi}=2 g m_B/f_{\pi}$.
The analogous coupling in charmed mesons can be measured
experimentally from the $D^*\to D\pi$ decay width, yielding 
$g=0.59\pm 0.01\pm 0.07$ 
\cite{Ahmed:2001xc,Anastassov:2001cw}.
Using this value in $B$ physics is subject to uncertainties
of order $\Lambda/m_c$.

At tree level, the coupling is related to the semi-leptonic
$B\to\pi l\nu$ decay form factor $f^+(q^2)$ in the large
$q^2$ region.
Using the soft pion theorem the form factor $f^+(q^2)$ near 
$q^2\sim m_{B^*}^2$ behaves as 
\begin{equation}
  f^+(q^2) = 
  \frac{f_{B^*}}{2m_{B^*}} 
  \frac{g_{B^*B\pi}}{(1-q^2/m_{B^*}^2)}.
\end{equation}
Thus, a precise knowledge of $g_{B^*B\pi}$ can constrain the
CKM matrix element $|V_{ub}|$.
The coupling $g$ also appears in the ChPT loop amplitudes as
in the discussion of chiral extrapolation of $f_B$ in
Section~\ref{sec:unquenched_BB}.

A direct lattice calculation of $g_{B^*B\pi}$ is possible if
one modifies Equation~\ref{eq:B*Bpi} by using the soft pion
theorem as a matrix element of the axial-vector current
\begin{equation}
  \langle B(p)|q_{\mu} A^{\mu}|B^*(p',\lambda)\rangle
  =
  g_{B^* B\pi} \frac{q\cdot\eta}{m_{\pi}^2-q^2}
  f_\pi m_\pi^2,
\end{equation}
where $q=p-p^{\prime}$. 
The lattice calculation of the matrix element on the left
hand side is technically similar to the calculation of
semi-leptonic form factors.
A quenched calculation in the static limit was done several
years ago and the result was $g=0.42\pm 0.04\pm 0.08$
\cite{deDivitiis:1998kj}.
More recently, Abada et al. performed an extensive
calculation for the $D^*D\pi$ coupling \cite{Abada:2002xe}
and for $B^*B\pi$ in the heavy quark mass limit
\cite{Abada:2003un}.
Their result $g=0.48\pm 0.03\pm 0.11$ \cite{Abada:2003un} is 
consistent with the previous estimates.
They also found that the $1/m_Q$ dependence is not
significant.

\subsection{Light-Cone Wave Function}
As discussed in Section~\ref{sec:Btopi}, $B$ decays with 
large recoil momentum are beyond the reach of current
lattice calculations.
Multi-hadron final states are even more difficult to treat
on the lattice.

For these energetic decays, a more natural theoretical
treatment is to factorize the energetic interaction, which
is perturbatively calculable, and the non-perturbative
physics, which is governed by the energy scale of order
$\Lambda_{\mathrm{QCD}}$. 
Such a formalism was developed by Beneke \textit{et al.} a
few years ago \cite{Beneke:1999br,Beneke:2001ev}.
For example, the amplitude of the non-leptonic $B\to\pi\pi$
decay is written as 
\begin{eqnarray}
  \langle \pi(p')\pi(q)| Q_i | B(p) \rangle 
  &=&
  f^{B\to\pi}(q^2) 
  \int_0^1\!\! du\, T_i^{\mathit{I}}(u) \Phi_{\pi}(u)
  \nonumber\\
  && + \int_0^1\!\! d\xi du dv  T_i^{\mathit{II}}(\xi,u,v) 
  \Phi_B(\xi)\Phi_{\pi}(u)\Phi_{\pi}(v).
\end{eqnarray}
Here $Q_i$ is a four-fermion operator and $T$'s are the 
perturbatively calculable short-distance part, whereas
$f^{B\to\pi}$ is the $B\to\pi l\nu$ semi-leptonic decay 
form factor, and $\Phi$'s are the light-cone wave functions.
Thus, the genuinely non-perturbative quantities are these
light-cone wave functions.

There are lattice studies on the light-cone wave function of
the pion. 
The light-cone wave function $\phi_{\pi}(u,Q^2)$ is defined
by the inverse Fourier transform of the matrix element 
\begin{equation}
  \langle 0 | \bar{d}(0) e^{\int_x^0 d\tau \cdot A(\tau)}
  \gamma_{\mu} \gamma_5 u(x) | \pi^+(p) \rangle |_{p^2=0} 
  =
  - ip_\mu f_\pi \int_0^1\!\! du\, e^{-ipx} \phi_{\pi}(u,Q^2). 
\end{equation}
The $n$-th moment of $\phi_{\pi}(u,Q^2)$, defined as 
$\langle \xi^n \rangle \equiv \int_0^1 d\xi \xi^n \phi_{\pi}(\xi,Q^2)$,
can be related to the pionic matrix element with the local
bilinear operator with $n$ derivatives as follows:
\begin{equation}
  \langle 0 | \bar{d} \gamma_{\mu} \gamma_5
  \stackrel{\leftrightarrow}{D}_{\mu_1} \cdots 
  \stackrel{\leftrightarrow}{D}_{\mu_n} u |   \pi^+(p) \rangle 
  =
  i^n f_\pi\, p_\mu p_{\mu_1} \cdots p_{\mu_n} \langle \xi^n \rangle.
\end{equation}
The second moment of the light-cone wave function has been
calculated by several lattice groups 
\cite{Gottlieb:ie,DeGrand:1987vy,Martinelli:1987si,%
  Martinelli:1987bh,Daniel:1990ah,DelDebbio:2002mq}.
A similar calculation is needed for the $B$ meson in order
to establish contact with the $B$ physics phenomenology.

A more direct calculation of the light-cone wave function,
not limited to its moments, has also been proposed
\cite{Aglietti:1998ur}.\footnote{
  This method can also be applied to the calculation of the
  $B$ meson shape function \cite{Aglietti:1998mz}, which
  describes the effect of the Fermi motion of $b$ quark on
  the inclusinve $B$ decays $B\to X_s\gamma$ and 
  $B\to X_ul\nu$. 
}
A pioneering numerical study has been carried out using this 
method \cite{Abada:2001if}, which is worthy of further
studies.


\section{Future perspectives}
\label{sec:Perspectives}

Lattice QCD has many applications, but the flavor physics 
is where it plays a crucial role in the exploration of
physics beyond the Standard Model through model
independent calculation of weak matrix elements.
Because the precision of the calculation is crucial in such 
studies, the systematic errors in the lattice calculation
have to be well understood and reduced as much as possible. 
In the past decade many important steps have been made
in this direction:
\begin{itemize}
\item 
  \textit{$O(a)$-improvement of lattice actions and operators.} 
  The discretization errors are described by Symanzik's
  effective theory, which also provides a method to improve
  lattice actions in a systematic way \cite{Symanzik:1983dc}.
  It is now common to use the $O(a)$-improved action, for
  which the leading discretization error is $O(a^2)$. 
  Further improvement is also being pursued.
\item 
  \textit{Effective theory for heavy quarks.}
  The ideas of HQET and NRQCD are naturally implemented on
  the lattice.
  The problem of dealing with heavy quarks on the lattice
  has essentially been solved
  \cite{Kronfeld:2002pi}.
\item 
  \textit{Improved perturbation theory.}
  Perturbation theory is still necessary in the lattice
  calculation to match the lattice action onto the target
  continuum theory.
  The convergence of the perturbative expansion was
  dramatically improved by the use of the renormalized
  coupling \cite{Lepage:1992xa}. 
\end{itemize}
The first two advances are related to the separation of
different energy scales by the use of effective theories,
and the third is needed for the matching of these effective 
theories.

We have reviewed the results that have relied on these 
methodological developments. 
One of the main challenges now is to further substantially
reduce the systematic error. 
This could be achieved by pushing the idea of the effective 
theories, \textit{i.e.} including the higher order
improvement terms and higher order relativistic
corrections. 
These have to be done together with the matching with the
higher order perturbation theory.
Although the higher order perturbative calculation is
technically rather demanding on the lattice, some work is 
already in progress \cite{Trottier:2003bw}.
Another proposed way to reduce the systematic errors, 
which does not rely on perturbative matching,
is the recursive non-perturbative matching
methods \cite{deDivitiis:2003wy,Heitger:2003nj}.

To obtain truly model independent results from lattice QCD, 
the quenched approximation must be abandoned.
Large scale simulations that include the effects of
dynamical quarks have already been carried out by several
lattice groups and will be performed more extensively in the 
future. 
One of the important effects of dynamical quarks is the pion 
loop effect, which leads to the chiral logarithm and affects
the chiral extrapolation of lattice data in a non-trivial
way.
As a result, the systematic uncertainty due to the chiral
extrapolation is still quite significant.
Although our discussion on this problem used $f_B$ as an
example, the same is true for almost all other lattice
observables as long as their chiral behavior is known from
ChPT. 
(Otherwise, there is no theoretical guide for the functional
form of chiral extrapolation.)
To eliminate the uncertainty from the chiral extrapolation,
one must push the dynamical quark masses in the lattice 
simulations down to the region where the chiral logarithm
becomes a visible effect.
It is presumably lower than $m_s/3$.
Such simulation is currently feasible only with the
staggered fermion action for dynamical quarks, which is
numerically so cheap that the MILC collaboration achieved
even $m_s/6$.
The results for several physical quantities agree with
their experimental values quite remarkably
\cite{Davies:2003ik}.
The price, however, is the introduction of fictitious
species, sometimes called ``tastes.''
The taste breaking effect introduces an additional source of
systematic uncertainty.
Furthermore, in the dynamical simulation the fourth root of
the fermionic determinant must be taken, 
and it is not known whether this gives a local field theory
in the continuum limit \cite{Jansen:2003nt}.
These points are still controversial.
Other lattice fermion formulations should also be pursued,
for which the developments of simulation algorithms are
essential.

As we have discussed, in order to achieve the goal of
precise, model-independent simulation of heavy quarks,
improvements are necessary throughout lattice QCD,
i.e. from the higher order perturbative calculation to the
simulation of really light dynamical quarks. 
Despite the difficulties, these future improvements will
be worth the effort, because from an improved flavor
physics one may probe physics beyond the Standard Model.

The application of lattice QCD to heavy quark physics is
not limited to $f_B$ and $B_B$.
Some other relevant quantities, such as semi-leptonic
decay form factors and heavy quark expansion parameters,
have already been studied on the lattice.
There are, however, many other important quantities that
require non-perturbative calculations.
The FCNC processes $B\to K^{(*)}l^+l^-$ are sensitive to 
new physics, and hence the calculation of their form factors
has direct relevance to the new-physics search.
The two-body decays of the $B$ meson have become
theoretically tractable since the factorization of short and
long distance physics was proven.
Non-perturbative inputs are necessary for the long distance
part, which is the light-cone distribution function.
Inclusive $B$ decays, such as $B\to X_s\gamma$ and 
$B\to X_ul\nu$, can be calculated perturbatively by using
the heavy quark expansion, but in the kinematical region of 
interest the simple OPE breaks down, so that resummation of
a certain class of higher-twist terms is necessary. 
Thus, a non-perturbative calculation of the quantity
called the shape function, which describes the Fermi motion 
of the bottom quark, is required to predict their energy
distributions.
Aside from the $B$ decays, the heavy meson and baryon
spectroscopy is also attracting attention, since the recent 
discovery of such exotic particles as $D_{sJ}(2317)$ and
$X(3872)$. 
Lattice QCD is the prime tool for the non-perturbative study
of these particles.
All these applications will be investigated in the future.


\section*{Acknowledgements}
We thank Andrew~G.~Akeroyd and Andreas~S.~Kronfeld for
carefully reading the manuscript and for their valuable
comments and discussions. 
The authors are supported in part by Grants-in-Aid for
Scientific Research from the Ministry of Education,
Culture, Sports, Science and Technology of Japan
(Nos. 13135213, 14540289, 16028210, 16540243).




\end{document}